\documentclass[onecolumn,showpacs,amsmath,amssymb]{revtex4}                 %
\usepackage{graphicx}
\begin{document}

\title{An investigation of a nonlocal entanglement of two uncoupled atoms embedded in a coherent cavity field and the associated phase space distribution: one quantum non-linear process}
\author{M. S. Ateto\footnote{E-mail: omersog@yahoo.com, mohammed.ali11@sci.svu.edu.eg}}
\affiliation{Mathematics Department, Faculty of Science at Qena, South Valley
University, 83523 Qena, Egypt
}
\date{\today}
\begin{abstract}
Entanglement properties of two uncoupled atoms embedded in a coherent field distribution
through one quantum transition process is studied. A case of non-linear Hamiltonian of the problem is considered through which the effect of a non-linear media is illustrated. Moreover, the effect of the frequency difference between the interatomic transition and the electromagnetic field is also analyzed. We show that, adjusting the considered parametres of the non-linear media and frequency difference leads to a strong control of the degree of entanglement where excellent periodicity of entanglement evolution can be obtained which is very important in predicting the behavior of transmitted information through the application of various information processing schemes.  We present a detailed and comparative study of atom-atom entanglement for two cases corresponding to different injections of the two atoms into the cavity field. Moreover, we present an answer to the question: How does the quantum phase space structure for a composite system relate to the entanglement characteristics of the corresponding quantum system? We demonstrate how the entanglement in nonlinear tripartite systems can be associated with a delocalization in the phase space distribution.

\end{abstract}
\maketitle
\section{Introduction}
\label{intro}

The recent rapid development of quantum information theory has largely stimulated research on nonclassical
phenomena, with the main focus on the generation of entangled states that are required for tasks such as quantum teleportation, dense coding, or certain types of quantum key distribution protocols. In recent years, there are much interest to study entanglement of two-level systems. These systems gain such  importance because of the efficiency in representing information in most quantum information processing schemes. Particular and great interest is devoted to the generation of entangled states in two-atom systems, since they can represent two qubits that considered the base blocks of building the quantum gates that are essential to implement various quantum protocols. Furthermore, there have been many proposals for generating atomic entanglement and entanglement between cavity modes through atom-photon interaction~ \cite{PWK04,BBPS96,BDSW96,VP96,V02,VT99,VPRK95,W98,GMN06,CZDKAW00,BGASMNN07,PK491,PK691,KLMCK93, ATETO109,Ateto07}. Some notable experimental demonstrations have also been performed~\cite{HMNWBRH979}.
\par However, the physical nature of the interacting objects and the
character of their mutual coupling control strongly the degree of quantum entanglement. A number of studies have shown that entanglement can be created between to objects which do not interact directly with each other but interact with either a common field or heat bath or thermal cavity field~\cite{BGASMNN07,ATETO109,Ateto07,KLAK02,BRAUN02, ATETO209}.
\par The formation of atom-photon entanglement and the subsequent generation of correlations between spatially separated atoms have been shown using the micromaser~\cite{MLBEHW96,ASMNN01,RSWRWR991,M02,HMNWBRH979}. The effectiveness of micromaser  ~\cite{FIJAME86,PK88,PB93} in generating entanglement has been appreciated as it is strongly considered as a practical device for processing information. It stores radiation for times significantly longer than the duration of the interaction with any single atom~\cite{PB93}. The interaction of an atom with the intracavity field of a micromaser will, generally, leave the atom-field system in an entangled state. The long cavity lifetime means that the memory of this entanglement can influence the interaction with subsequent atoms and a nonlocal correlations between these successive atoms can be induced leading to a violation of Bell's inequality~\cite{PB93}.
\par To our knowledge, the study of such a problem when the cavity field contains a coherent distribution is still insufficient. The major difficulty may be due to the complex calculations and long analytical mathematics, specially when the system Hamiltonian becomes nonlinear. We are interested in the opportunity to use the micromaser with coherent distribution as a source of nonlocal entanglement and answer the important questions: how and to what extent can the degree of atom-atom entanglement be cotrolled in these circumstances? Also, can the Husimi $Q$-distribution be used as a helpful method to quantify nonlocal entanglement dynamics? When applied to a system of two qubits, it’s shown that the concurrence~\cite{W98} may be expressed in terms of the second moment of the Husimi $Q$-function. More strikingly, when applied to a system of three qubits, the expression for the moment contains all three bipartite concurrence terms. In this way this measure captures all classes of entanglement, not only bipartite or otherwise~\cite{SUGITA03}.
\par Our purpose here is to trace analytically and numerically the problem of a nonlocal entanglement created between two successive atoms (two-qubits) through surrounding the coherent cavity field with a nonlinear medium, namely,  Kerr-like medium. The Kerr medium~\cite{FALI95, ATETO109, ATETO209, ATETO10, Ateto07, WLG06} can be modelled as an anharmonic oscillator with frequency $\omega$. Physically this model may be realized as if the cavity contains two different species of atoms, one of which behaves like a two-level atom and the other behaves like an anharmonic oscillator in the single-mode field of frequency $\omega$~\cite{JOPU92}. Such a model is interesting by itself. The cavity mode is coupled to the Kerr medium as well as to the two-level atoms. A Kerr-like medium can be useful in many respects, such as detection of nonclassical states~\cite{Hillaer91}, quantum nondemolition measurement~\cite{CHCOWO92}, investigation of quantum fluctuations~\cite{ZHGCLM00}, generation of entangled macroscopic quantum states\cite{Gerry99,PKH03}, and quantum information processing~\cite{PACH00,VFT00}.
\par An outline of paper is as follows: In Sec. II we describe in details the physical mechanism of the interaction and introduce the modified model of the total system that accommodates processes such as nonlocal interaction in nonlinear media. We obtain the wave function of the first part of interaction on which the final wave function of the total system depends. In Sec. III, we describe the technique we are going to use in quantifying the degree of entanglement. In addition, we calculate the the eigenvalues of the non-Hermitian operator that used to compute the mathematical form of entanglement measure. Different kinds of atoms injection into the cavity are introduced in sections IV and V, where the corresponding wave functions and their associated density matrix operators are computed. In Set. VI, we give a little historical review about the Husimi $Q$-function, where that function is calculated for different circumstances separately with detailed discusion included. Our concluding remarks are shown in Set. VII.
\section{The full system and its solution}
\label{sec:2}
\par We consider two two-level atoms with the same transition frequency traverse a high $Q$ ($Q\approx10^{9}$) single mode cavity one after the other. The flight time of the atoms is, however, long enough so that there is no appreciable overlap between the atomic wave functions~\cite{BGASMNN07,DGMN04,LEW96,AMNN01,PB93,Ateto07,EGSWH96,MHNWBTH97,ELSW002}. With this assumption we ignore the possibility of energy exchange between the atoms, although, secondary correlations develop between them. The entanglement of their wave functions with the cavity photons can be used to formulate local-realist bounds on the detection probabilities for the two atoms~\cite{LEW96,PB93}. The
generation of nonlocal correlations between the two atomic states emerging from the cavity can in general
be understood using the Horodecki theorem~\cite{HHH95}. It is assumed that the atom field interaction time is shorter than the lifetime of the cavity, so that the cavity relaxation will not be considered. The cavity field is assumed to be filled with a nonlinear medium, namely, Kerr medium~\cite{FALI95,ATETO109,ATETO10, ATETO209, Ateto07,WLG06}. Also, it is assumed that the eigenfrequency of the atomic subsystem differs from that of cavity field subsystem. Under the rotating wave approximation (TWA), the Hamiltonian in the interaction picture (assuming that $\hbar=1$)
\begin{equation}
\label{eq1}
\widehat{H}_{int}=\frac{\Delta}{2}~\widehat{\sigma}_{z}+f(\chi,
\widehat{n})
+\lambda~\bigl(\widehat{\sigma}_{-}~\widehat{a}^{\dag}+\widehat{\sigma}_{+}~\widehat{a}\bigr),
\end{equation}
where $f(\chi, \widehat{n})=\chi  (\widehat{n}^{2}-\widehat{n})$ represents the nonlinear term with,
\begin{equation}
\label{eq2}
f(\chi, \widehat{n})\mid n\rangle=\bigl[\chi
(n^{2}-n)\bigr]\mid n\rangle=f(\chi, n)\mid n\rangle,
\end{equation}
where $\widehat{a}$ ($\widehat{a}^{\dag}$) is bosonic annihilation (creation) operator for the single mode field of frequency $\omega$ and $\lambda$ is atoms-field coupling constant, the operators $\widehat{\sigma}_{+}=| +\rangle\langle -|$ , $\widehat{\sigma}_{-}=|-\rangle\langle +|$ and $\widehat{\sigma}_{z}=|+\rangle\langle +|-|-\rangle\langle-|$  
represent, respectively, the raising, lowering and population atomic operators which obey the commutation
relations $[\widehat{\sigma}_{+},\widehat{\sigma}_{-}]=\widehat{\sigma}_{z}$,
while $\Delta=\omega_{0}-\omega$ is the detuning parameter which represents the difference between the atomic and cavity subsystems eigenfrequencies. We denote by $\chi$ the dispersive part of the third order susceptibility of the Kerr-like medium~\cite{FALI95, ATETO109,ATETO10, ATETO209, Ateto07, WLG06}.\\
Since the state vector, $|\psi_{F}(0)\rangle$, of the field is represented by a linear superposition of the number state $|n\rangle$
\begin{equation}
\label{eq3}
|\psi_{F}(0)\rangle=\sum_{n=0}^{\infty} C^{n}|n\rangle, 
\end{equation}
and the first atom that traverse the cavity is assumed in its upper state $\mid +\rangle$, i. e.,
\begin{equation}
\label{eq4}
|\psi_{A}(0)\rangle=\mid +\rangle
\end{equation}
the initial state vector of
the interacting first-atom-field system is given by
\begin{equation}
\label{eq5}
|\psi_{AF}(0)\rangle=|\psi_{A}(0)\rangle \otimes
|\psi_{F}(0)\rangle=\sum_{n=0}^{\infty} C^{n}|n, +\rangle,
\end{equation}
where $|n\rangle$ is an eigenstate of the number operator $\widehat{a}^{\dagger}\widehat{a}=\widehat{n}$%
; $\widehat{a}^{\dagger}\widehat{a} |n\rangle=n|n\rangle$, and $C^{n}$ is, in general, complex where the square of its modulus gives the probability of the coherent field, with mean photon $|\alpha|^{2}=\bar{n}$, to have $n$ photons by the relation
\begin{equation}
\label{eq6}
P(n)=|\langle n|\psi_{F}(0)\rangle|^2=|C^{n}|^{2}=e^{-\bar{n}}\frac{\bar{n}^{n}}{n!}.
\end{equation}
Under the assumption that, first atom traverse the cavity in its upper state $\mid +\rangle$, the joint state vector of the
field and the first atom at any instant of time $t$ can be obtained from the solution of the time-dependent  Schr\"{o}dinger equation
\begin{equation}
\label{eq7}
i \frac {d}{dt}|\psi_{AF}(t)\rangle=\widehat{H}~|\psi_{AF}(t)\rangle.
\end{equation}
The time-dependent wave function of the first-atom-field system takes the form
\begin{equation}
\label{eq8}
|\psi_{AF}(t)\rangle=\mid U_{+}^{n}(t)\rangle+\mid U_{-}^{n+1}(t)\rangle
\end{equation}
which can be simply obtained by recalling the initial condition (\ref{eq5}) and solving the Schr\"{o}dinger equation (\ref{eq7}), where
\begin{equation}
\label{eq9}
\mid U_{+}^{n}(t)\rangle=\sum_{n}^{\infty}~C^{n}~ \Gamma_{1}(n,t)~|n,+
\rangle,
\end{equation}
\begin{equation}
\label{eq10}
\mid U_{-}^{n+1}(t)\rangle=\sum_{n}^{\infty}~C^{n} ~\Gamma_{2}(n+1,t)~|n+1,-\rangle,
\end{equation}
with the amplitudes $\Gamma_{1}(n,t)$ and $\Gamma_{2}(n+1,t)$ are, respectively, given by
\begin{equation}
\label{eq11}
\Gamma_{1}(n,t)=\sum_{j=1}^{2}F_{j}(n) e^{i\mu_{j}(n) t},
\end{equation}
and
\begin{equation}
\label{eq12}
\Gamma_{2}(n+1,t)=-\sum_{j=1}^{2}F_{j}\frac{\mu_{j}(n)+\alpha(n)}{\lambda\sqrt{n+1}} e^{i\mu_{j}(n) t},
\end{equation}
with $F_{j}(n)$ is given by 
\begin{equation}
\label{eq13}
F_{j}=\frac{\mu_{k}(n)+\alpha(n)}{\mu_{kj}(n)},~~~~k\neq j=1,2,
\end{equation}
where the angles $\mu_{1,2}(n)$ are given by
\begin{equation}
\label{eq14}
\mu_{1,2}(n)=-\frac{\alpha(n)+\gamma(n)}{2}\pm\sqrt{\biggl(\frac{\alpha(n)-\gamma(n)}{2}\biggr)^{2}+\lambda^{2}(n+1)}.
\end{equation}
with $\alpha(n)$ and $\gamma(n)$ read, respectively
\begin{equation}
\label{eq15}
\alpha(n)=\frac{\Delta}{2}+\chi n(n-1),
\end{equation}
\begin{equation}
\label{eq16}
\gamma(n)=-\frac{\Delta}{2}+\chi n(n+1).
\end{equation}
\par As already indicated above, we consider a pair of two-level atoms going
through the cavity mode one after another. Moreover, under the assumption that time of flight through the cavity $t$ is the same for every
atom~\cite{BGASMNN07,DGMN04,LEW96,AMNN01,PB93,Ateto07,EGSWH96,MHNWBTH97,ELSW002}, where the joint state vector of both atoms and the field may be denoted by $|\psi_{AAF}(t)\rangle$, the corresponding atom-atom-field pure-state density operator may be written as
\begin{equation}
\label{eq17}
\rho_{AAF}(t)=|\psi_{AAF}(t)\rangle \langle\psi_{AAF}(t)|,
\end{equation}
where, the joint time-evolved wave vector, $|\psi_{AAF}(t)\rangle$, of the tripartite system, i.e., the two atoms and the cavity field after the second atom leaves the cavity is obtained by solving the Schr\"{o}dinger equation
\begin{equation}
\label{eq18}
i\frac {d}{dt}|\psi_{AAF}(t)\rangle=\widehat{H}_{int}|\psi_{AAF}(t)\rangle.
\end{equation}
It is worth to note that within the delay time between the two atoms the field evolves towards a thermal steady state, moreover, repetition of the instant in which the later atoms enter the cavity means the same field repeats at this instants precisely when successive atoms exit the cavity~\cite{FIJAME86}.\\
In order to quantify the degree of entanglement between the two atoms, the
field variables must be traced out. One may write the reduced mixed-state density
matrix of the two atoms after taking the trace over the field variables as
\begin{equation}
\label{eq19}
\mathbf{\rho_{AA}}(t)=\mathrm{Tr}_{F}~[\rho_{AAF}(t)].
\end{equation}
Matrix representation of the reduced density operator, Eq. (\ref{eq19}) becomes
\begin{equation}
\label{eq20}
\mathbf{\rho_{AA}}(t) = \left(
\begin{array}{cccc}
\rho_{11} & \rho_{12} & \rho_{13} & \rho_{14} \\
 \rho_{21}& \rho_{22} & \rho_{23} & \rho_{24} \\
\rho_{31} & \rho_{32} & \rho_{33} & \rho_{34} \\
\rho_{41} & \rho_{42} & \rho_{43} & \rho_{44}%
\end{array}
\right),
\end{equation}
\section{Entanglement measure}
\label{sec:3}
For bipartite pure states, the partial (von Neumann) entropy of the reduced density matrices can provide a good measure of entanglement. However, for mixed states von Neumann entropy fails, because it can not distinguish classical and quantum mechanical correlations. For mixed states, the entanglement can be measured as the average entanglement of its pure-state decompositions $%
E_{F}(\rho)$
\begin{equation}
\label{eq21}
E_{F}(\rho)=min \sum_{i} p_{i} E(\psi_{i}),
\end{equation}
with $E(\psi_{i})$ being the entanglement measure for the pure state $\psi_{i}$
corresponding to all the possible decompositions $\rho=\sum_{i}
p_{i}|\psi_{i}\rangle\langle\psi_{i}|$. The existence of an infinite number
of decompositions makes their minimization over this set difficult.
Wootters~\cite{W98} succeeded in deriving an analytical solution to this difficult
minimization procedure in terms of the eigenvalues of the non-Hermitian
operator
\begin{equation}
\label{eq22}
T=\rho \tilde{\rho},
\end{equation}
where the tilde denotes the spin-flip of the quantum state, which is defined
as
\begin{equation}
\label{eq23}
\tilde{\rho}=(\sigma_{x}\otimes\sigma_{x})\rho^{\ast}(\sigma_{x}\otimes%
\sigma_{x}),
\end{equation}
where $\sigma_{x}$ is the Pauli matrix, and $\rho^{\ast}$ is the complex
conjugate of $\rho$ where both are expressed in a fixed basis such as $%
\{|+\rangle, |-\rangle\}$.
\\
In terms of the eigenvalues of $T=\rho \tilde{\rho}$, $E_{F}(\rho)$ (known
as the entanglement of formation) takes the form
\begin{equation}
\label{eq24}
E_{F}(\rho)=h\biggl[\frac{1}{2}+\frac{1}{2}\sqrt{1-C^{2}(\rho)}\biggr],
\end{equation}
where $C(\rho)$ is called the concurrence and is defined as
\begin{equation}
\label{eq25}
C(\rho)=\max\biggl(0, \sqrt{\mathcal{E}_{1}}-\sqrt{\mathcal{E}_{2}}-\sqrt{\mathcal{E}_{3}}-%
\sqrt{\mathcal{E}_{4}}\biggr),
\end{equation}
with the $\mathcal{E}$'s representing the eigenvalues of $T=\rho \tilde{\rho}$ in
descending order, and,
\begin{equation}
\label{eq26}
h(y)=-y \log y-(1-y) \log (1-y),
\end{equation}
is the binary entropy. The concurrence is associated with the entanglement of
formation $E_{F}(\rho)$, Eq.(\ref{eq24}), but it is by itself a good measure for
entanglement. The range of concurrence is from 0 to 1. For unentangled atoms
$C(\rho)=0$ whereas $C(\rho)=1$ for maximally entangled atoms.\\
Recalling (\ref{eq23}) and assuming that that the matrix $T=\rho_{AA}(\sigma_{x}\otimes \sigma_{x})\rho_{AA}^{\ast}(\sigma_{x}\otimes \sigma_{x})$, needed for calculation of the concurrence, has the form
\begin{equation}
\label{eq27}
T = \left(
\begin{array}{cccc}
T_{11} & T_{12} & T_{13} & T_{14} \\
 T_{21}& T_{22} & T_{23} & T_{24} \\
T_{31} & T_{32} & T_{33} & T_{34} \\
T_{41} & T_{42} & T_{43} & T_{44}%
\end{array}
\right),
\end{equation}
the next step is to find the eigenvalues of the above matrix $T$. To achieve our goal, we are supposed to solve the
characteristic equation of
\begin{equation}
\label{eq28}
\mathrm{Det}(T- \mathcal{E}\mathbf{I})=0,
\end{equation}
Equation (\ref{eq28}) is a polynomial equation of degree 4
\begin{equation}
\label{eq29}
{\mathcal{E}}^4 +c_3 {\mathcal{E}}^3 +c_2 {\mathcal{E}}^2 +c_1 {\mathcal{E}}+c_0
=0,
\end{equation}
with
\begin{equation}
\label{eq30}
c_{3}=-T_{11}-T_{22}-T_{33}-T_{44},
\end{equation}
\begin{equation}
\label{eq31}
c_{2}=-|T_{13}|^{2}-|T_{14}|^{2}-|T_{12}|^{2}-|T_{34}|^{2}-|T_{24}|^{2}
+(T_{11}+T_{22})(T_{33}+T_{44})+T_{11}T_{22}+T_{33}T_{44},
\end{equation}
\begin{equation}
c_{1}=|T_{12}|^{2}(T_{33}+T_{44})+|T_{23}|^{2}(T_{11}+T_{44})
|T_{24}|^{2}(T_{11}+T_{33})+|T_{34}|^{2}(T_{11}+T_{22})
|T_{13}|^{2}(T_{22}+T_{44})+|T_{14}|^{2}(T_{22}+T_{33})\nonumber
\end{equation}
\begin{equation}
\label{eq32}
-\Re(T_{23}T_{34}T_{42})-\Re(T_{21}T_{32}T_{13})
-\Re(T_{21}T_{42}T_{14})-\Re(T_{31}T_{14}T_{43})
-T_{11}(T_{11}T_{33}+T_{22}T_{33}+T_{11}T_{22}),
\end{equation}
\begin{equation}
c_{0}=T_{11}T_{22}T_{33}T_{44}+|T_{13}|^{2}|T_{24}|^{2}+|T_{14}|^{2}|T_{23}|^{2}
-T_{11}T_{22}|T_{34}|^{2}-T_{11}T_{44}|T_{23}|^{2}-T_{33}T_{44}|T_{12}|^{2}\nonumber
\end{equation}
\begin{equation}
-T_{11}T_{33}|T_{24}|^{2}-T_{22}T_{33}|T_{14}|^{2}
+T_{11}\Re(T_{32}T_{24}T_{43})+T_{22}\Re(T_{31}T_{14}T_{43})
+T_{33}\Re(T_{21}T_{42}T_{14})+T_{44}\Re(T_{32}T_{13}T_{43})\nonumber
\end{equation}
\begin{equation}
\label{eq33}
-\Re(T_{21}T_{32}T_{14}T_{43})-\Re(T_{21}T_{42}T_{13}T_{34})
-\Re(T_{31}T_{42}T_{14}T_{23}).
\end{equation}
The roots of Eq.(\ref{eq29}) are as follows
\begin{equation}
\label{eq34}
\mathcal{E}_{1}=-\frac{1}{4}\biggl[c_{3}-\sqrt{\frac{W}{3}}-4 O\biggr]
\end{equation}
\begin{equation}
\label{eq35}
\mathcal{E}_{2}=-\frac{1}{4}\biggl[c_{3}-\sqrt{\frac{W}{3}}+4 O\biggr]
\end{equation}
\begin{equation}
\label{eq36}
\mathcal{E}_{3}=-\frac{1}{4}\biggl[c_{3}+\sqrt{\frac{W}{3}}-4 N\biggr]
\end{equation}
\begin{equation}
\label{eq37}
\mathcal{E}_{4}=-\frac{1}{4}\biggl[c_{3}+\sqrt{\frac{W}{3}}+4 N\biggr]
\end{equation}
with
\begin{equation}
\label{eq38}
O=\frac{1}{2}\sqrt{\frac{O_{1}+O_{2}+O_{3}}{6}}
\end{equation}
\begin{equation}
\label{eq38}
N=\frac{1}{2}\sqrt{\frac{O_{1}+O_{2}-O_{3}}{6}}
\end{equation}
where
\begin{equation}
\label{eq39}
O_{1}=3c_{3}^{2}-8c_{2}-V^{1/3}
\end{equation}
\begin{equation}
\label{eq40}
O_{2}=4 V^{-1/3}[3 c_{1}c_{3}-12 c_{0}-c_{2}^{2}]
\end{equation}
\begin{equation}
\label{eq41}
O_{3}=\sqrt{\frac{3}{w}}[4 c_{3}c_{2}-c_{3}^{3}-8 c_{1}]
\end{equation}
and
\begin{equation}
\label{eq42}
V=V_{1}+\sqrt{\frac{V_{2}+V_{3}+V_{4}+V_{5}}{144}}
\end{equation}
with
\begin{equation}
\label{eq43}
V_{1}=8c_{2}^{2}+36[3(c_{1}^{2}+c_{0}c_{3}^{2})-c_{1}c_{2}c_{3}-c_{0}c_{2}]
\end{equation}
\begin{equation}
\label{eq44}
V_{2}=-54c_{2}[c_{0}c_{1}c_{3}^{3}+c_{1}^{3}c_{3}+8c_{0}(c_{1}^{2}+c_{0}c_{3}^{2})]
\end{equation}
\begin{equation}
\label{eq45}
V_{3}=3c_{1}c_{3}[2c_{0}(4c_{2}^{2}+3c_{1}c_{3})-c_{1}c_{2}^{2}c_{3}]
\end{equation}
\begin{equation}
\label{eq46}
V_{4}=12[(c_{1}c_{3})^{3}+4c_{0}c_{2}^{2}(8c_{0}-c_{2}^{2})+c_{2}^{3}(c_{1}^{2}+c_{0}c_{3}^{2})]
\end{equation}
\begin{equation}
\label{eq47}
V_{5}=c_{0}^{2}[81c_{3}^{4}+576c_{1}c_{3}-768c_{c}]
\end{equation}
and
\begin{equation}
\label{eq48}
W=W_{1}+W_{2}
\end{equation}
where
\begin{equation}
\label{eq49}
W_{1}=3c_{3}^{2}-8c_{2}+2V^{1/3}
\end{equation}
\begin{equation}
\label{eq50}
W_{2}=8 V^{-1/3}[c_{2}^{2}+3(4c_{0}-c_{1}c_{3})]
\end{equation}
\par In the following we consider two different cases of injection of the second atom into the cavity subsystem that give us clear insight into the mechanism of how and to what extent the degree of entanglement can be controlled and enhanced.
\section{injection of two excited atoms successfully}
\label{sec:4}
In this case, the time dependent state vector of the full system can be expressed in the form
\begin{equation}
\label{eq52}
|\psi_{AAF}(t)\rangle=\biggl| U_{\begin{array}{l}
\vspace{-2ex}+\\
             +\\
\end{array}
}(t)\biggr >\mid n\rangle+\biggl| U_{\begin{array}{l}
\vspace{-2ex}+\\
             -\\
\end{array}
}(t)\biggr >\mid n+1\rangle+\biggl| U_{\begin{array}{l}
\vspace{-2ex}-\\
             +\\
\end{array}
}(t)\biggr >\mid n+1\rangle+\biggl| U_{\begin{array}{l}
\vspace{-2ex}-\\
             -\\
\end{array}
}(t)\biggr >\mid n+2\rangle,
\end{equation}
which can be simply obtained by solving the Shr{\"o}dinger equation (\ref{eq18}), where
\begin{equation}
\label{eq53}
\biggl| U_{\begin{array}{l}
\vspace{-2ex}+\\
             +\\
\end{array}
}(t)\biggr >=\sum_{n} C^{n}~[\Gamma_{1}(n,t)]^{2}\bigl|\begin{array}{l}
\vspace{-2ex}+\\
             +\\
\end{array}\bigr>,
\end{equation}
\begin{equation}
\label{eq54}
\biggl| U_{\begin{array}{l}
\vspace{-2ex}+\\
             -\\
\end{array}
}(t)\biggr >=\sum_{n} C^{n}~\Gamma_{1}(n,t)\Gamma_{2}(n+1,t)\bigl|\begin{array}{l}
\vspace{-2ex}+\\
             -\\
\end{array}\bigr>,
\end{equation}
\begin{equation}
\label{eq55}
\biggl| U_{\begin{array}{l}
\vspace{-2ex}-\\
             +\\
\end{array}
}(t)\biggr >=\sum_{n} C^{n}~\Gamma_{1}(n+1,t)\Gamma_{2}(n+1,t)\bigl|\begin{array}{l}
\vspace{-2ex}-\\
             +\\
\end{array}\bigr>,
\end{equation}
\begin{equation}
\label{eq56}
\biggl| U_{\begin{array}{l}
\vspace{-2ex}-\\
             -\\
\end{array}
}(t)\biggr >=\sum_{n} C^{n}~\Gamma_{2}(n+1,t)\Gamma_{2}(n+2,t)\bigl|\begin{array}{l}
\vspace{-2ex}-\\
             -\\
\end{array}\bigr>,
\end{equation}
Using (\ref{eq17}) to obtain the full density operator and after the application of (\ref{eq19}) to obtain the reduced atomic density operator which can be put, after using the notations
$\bigl|\begin{array}{l}
\vspace{-2ex}+\\
             +\\
\end{array}\bigr>\equiv\mid 1\rangle$, $\bigl|\begin{array}{l}
\vspace{-2ex}+\\
             -\\
\end{array}\bigr>\equiv\mid 2\rangle$, $\bigl|\begin{array}{l}
\vspace{-2ex}-\\
             +\\
\end{array}\bigr>\equiv\mid 3\rangle$, $\bigl|\begin{array}{l}
\vspace{-2ex}-\\
             -\\
\end{array}\bigr>\equiv\mid 4\rangle$, in the form
\begin{equation}
\label{eq57}
\rho_{AA}(t)=\sum_{i,j=1}^{4}~\rho_{ij}(t)|i\rangle\langle j|,
\end{equation}
where
\begin{equation}
\label{eq58}
\rho_{11}(t)=\sum_{n}\bigl|C^{n}\bigr|^{2}\bigl|\Gamma_{1}(n,t)\bigr|^{4},
\end{equation}

\begin{equation}
\label{eq59}
\rho_{12}(t)=\sum_{n} C^{n+1}C^{\ast n}\bigl[\Gamma_{1}(n+1,t)\bigr]^{2}\Gamma_{1}^{\ast}(n,t)\Gamma_{2}^{\ast}(n+1,t),
\end{equation}
\begin{equation}
\label{eq60}
\rho_{13}(t)=\sum_{n} C^{n+1}C^{\ast n}\bigl|\Gamma_{1}(n+1,t)\bigr|^{2}\Gamma_{1}(n+1,t)\Gamma_{2}^{\ast}(n+1,t),
\end{equation}
\begin{equation}
\label{eq61}
\rho_{14}(t)=\sum_{n} C^{n+2}C^{\ast n} \bigl[\Gamma_{1}(n+2,t)\bigr]^{2}\Gamma_{2}^{\ast}(n+1,t)\Gamma_{2}^{\ast}(n+2,t),
\end{equation}
\begin{equation}
\label{eq62}
\rho_{22}(t)=\sum_{n} \bigl|C^{n}\bigr|^{2}\bigl|\Gamma_{1}(n,t)\bigr|^{2}\bigl|\Gamma_{2}(n+1,t)\bigr|^{2},
\end{equation}
\begin{equation}
\label{eq63}
\rho_{23}(t)=\sum_{n} \bigl|C^{n}\bigr|^{2} \bigl|\Gamma_{2}(n+1,t)\bigr|^{2}\Gamma_{1}(n,t)\Gamma_{1}^{\ast}(n+1,t),
\end{equation}
\begin{equation}
\label{eq64}
\rho_{24}(t)=\sum_{n} C^{n+2}C^{\ast n} \Gamma_{1}(n+1,t)\Gamma_{2}^{\ast}(n+1,t)\bigl|\Gamma_{2}(n+2,t)\bigr|^{2},
\end{equation}
\begin{equation}
\label{eq65}
\rho_{33}(t)=\sum_{n} \bigl|C^{n}\bigr|^{2}\bigl|\Gamma_{1}(n+1,t)\bigr|^{2}\bigl|\Gamma_{2}(n+1,t)\bigr|^{2},
\end{equation}
\begin{equation}
\label{eq66}
\rho_{34}(t)=\sum_{n} C^{n+2}C^{\ast n} \Gamma_{1}(n+1,t)\Gamma_{2}^{\ast}(n+1,t)\bigl|\Gamma_{2}(n+2,t)\bigr|^{2},
\end{equation}
\begin{equation}
\label{eq67}
\rho_{44}(t)=\sum_{n} \bigl|C^{n}\bigr|^{2}\bigl|\Gamma_{2}(n+1,t)\bigr|^{2}\bigl|\Gamma_{2}(n+2,t)\bigr|^{2},
\end{equation}
For $\chi=\Delta=0.0$, the results reported in Ref. \cite{BGASMNN07} are straightforward obtained, which analyzed in details but for low and high average photon number $\bar n$. \\
\begin{figure}[tpbh]
\noindent
\begin{center}
\includegraphics[width=.45\linewidth]
{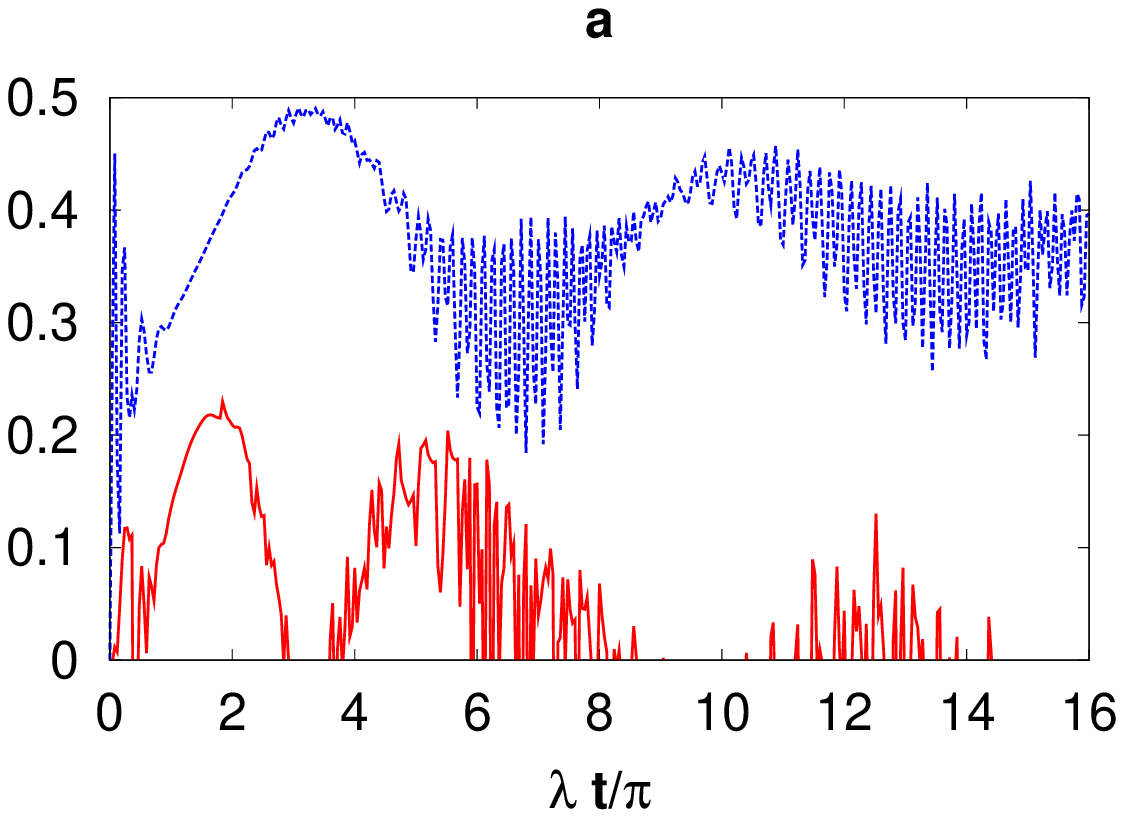}
\includegraphics[width=.45\linewidth]
{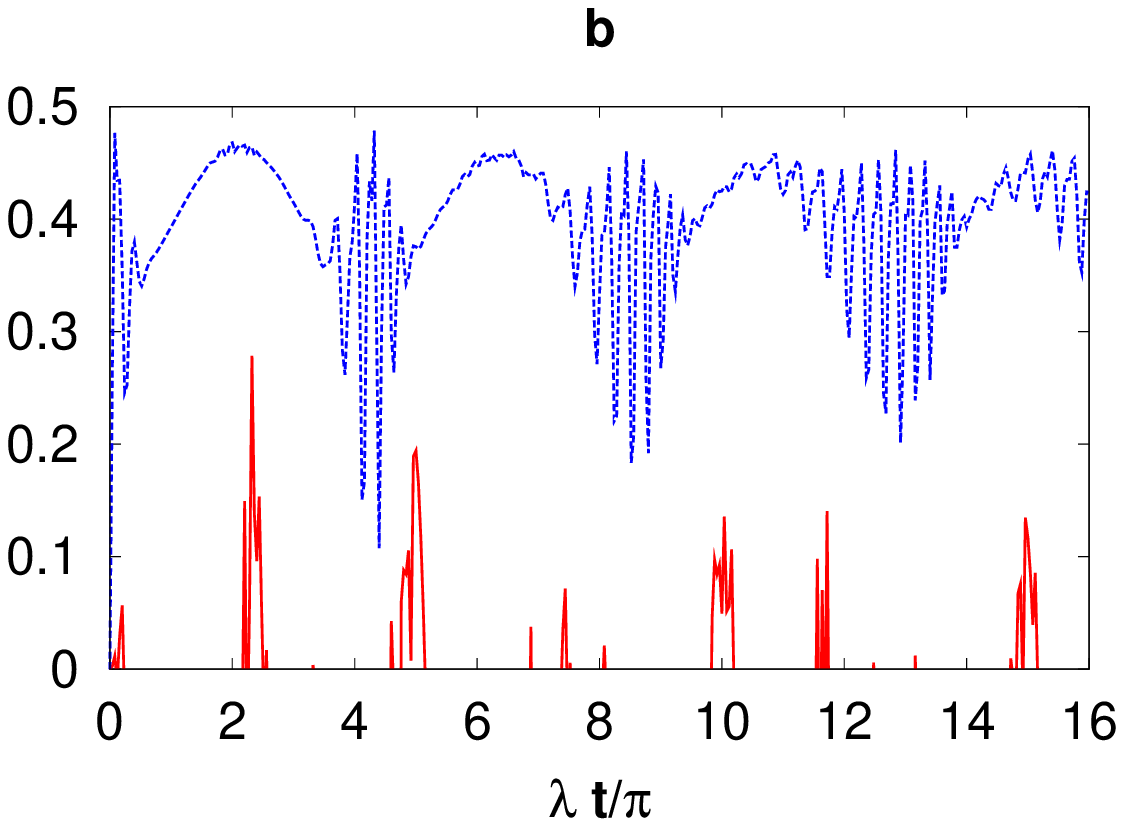}
\includegraphics[width=.45\linewidth]
{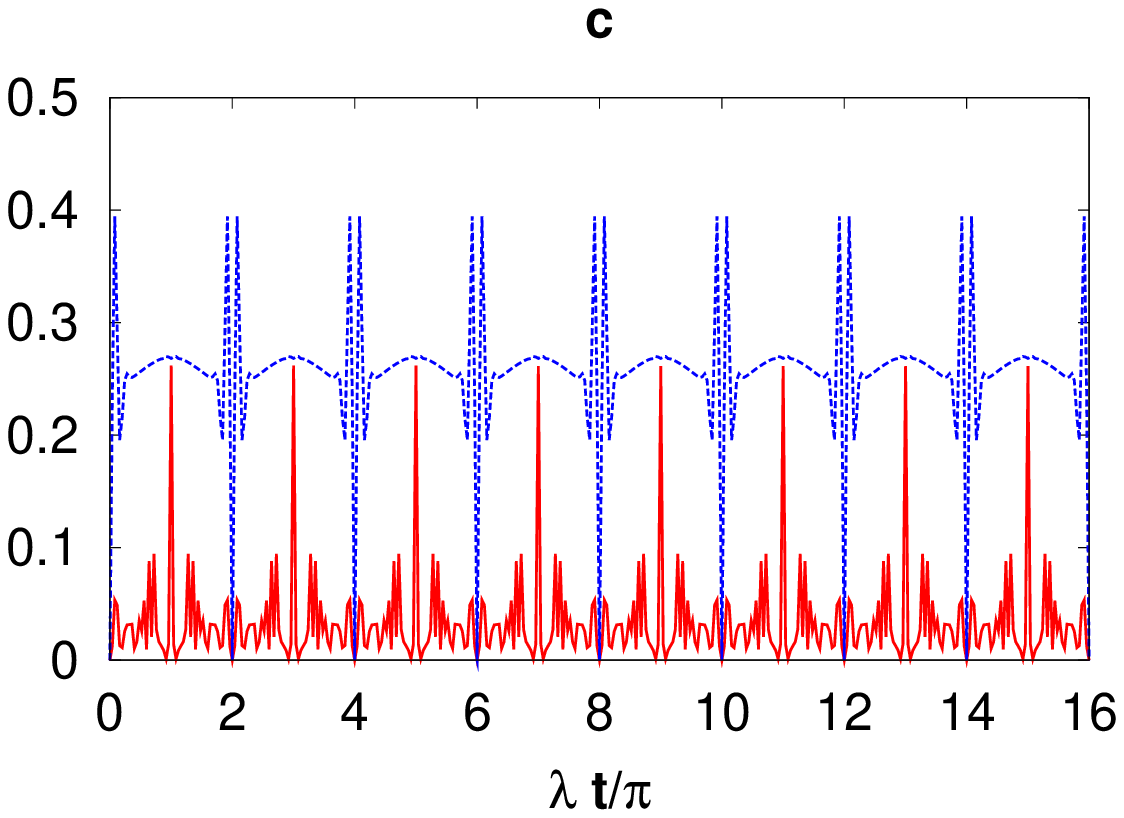}
\includegraphics[width=.45\linewidth]
{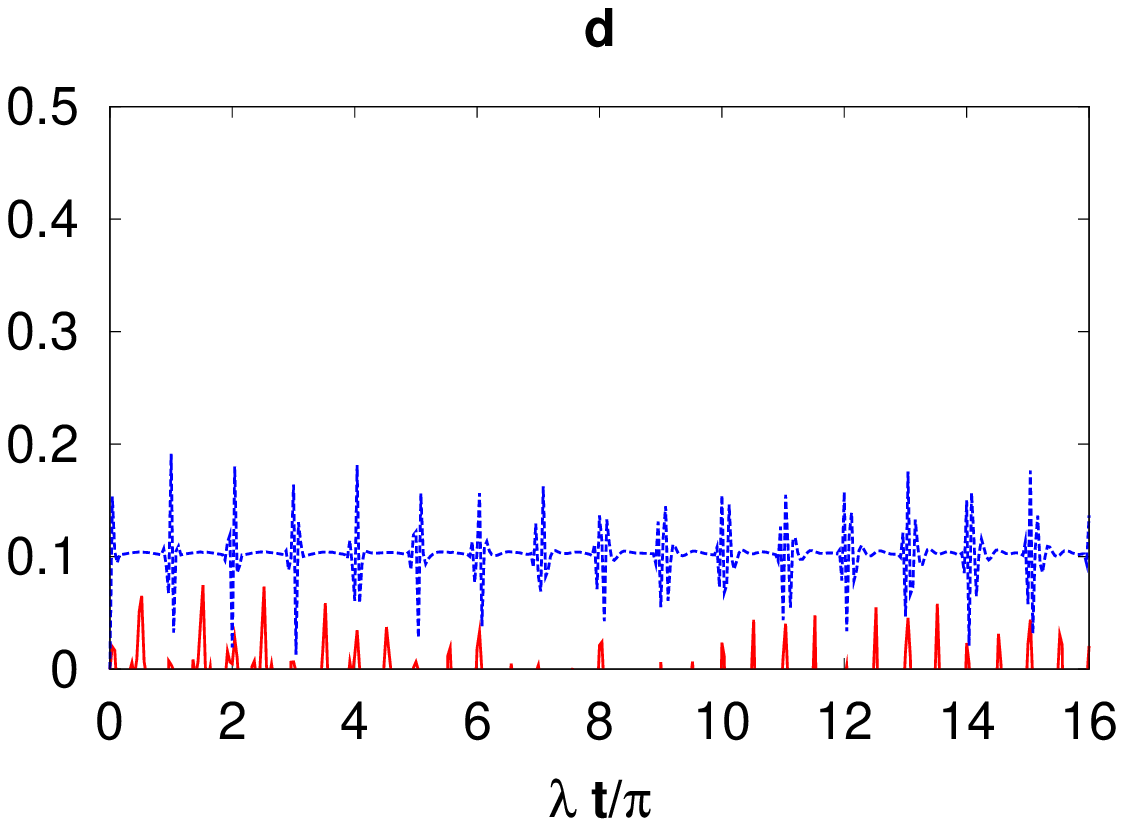}
\end{center}
\caption{The concurrence $C(\lambda t/\pi)$ (solid red curve) and the total population $\rho_{22}(\lambda t/\pi)+\rho_{33}(\lambda t/\pi)$ (dashed blue curve) for $\bar{n}=10$, $\delta=0.0$ where (a) $\chi/\lambda=0.0$ (b) $\chi/\lambda=0.2$ (c) $\chi/\lambda=0.5$ (c) $\chi/\lambda=1.0$ }
\end{figure}
\begin{figure}[tpbh]
\noindent
\begin{center}
\includegraphics[width=.45\linewidth]
{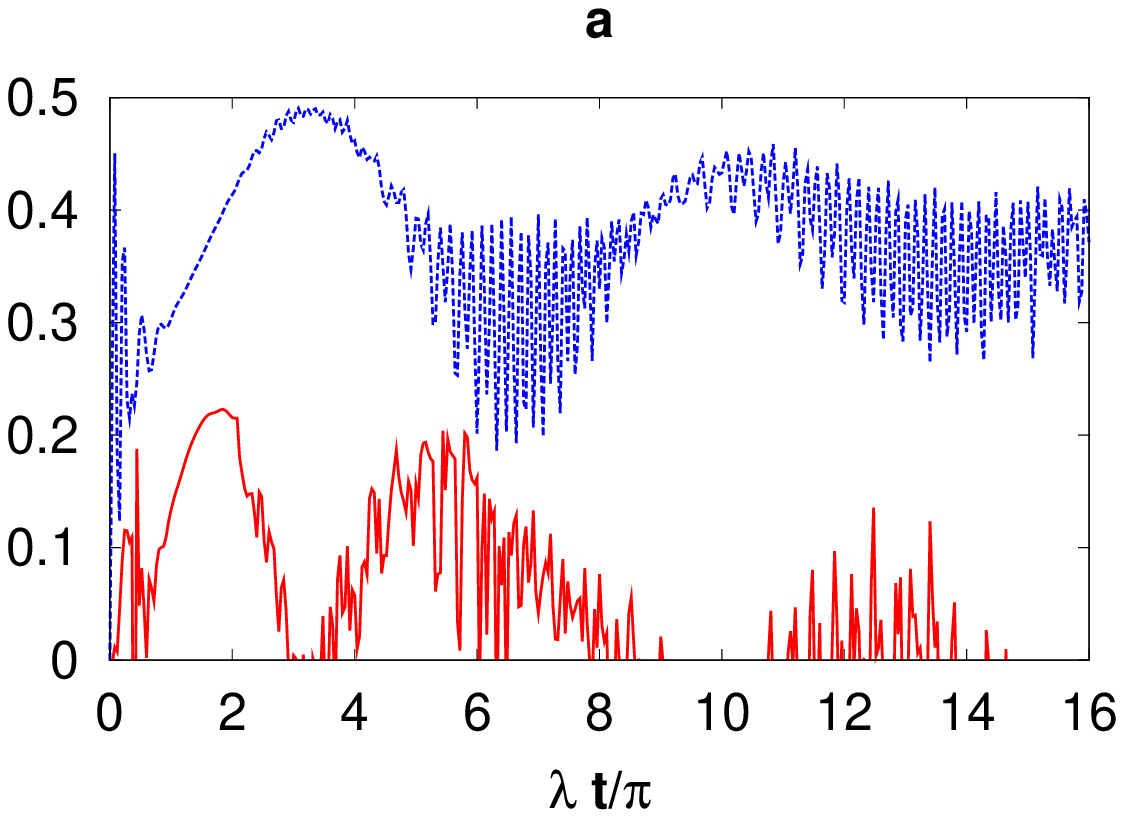}
\includegraphics[width=.45\linewidth]
{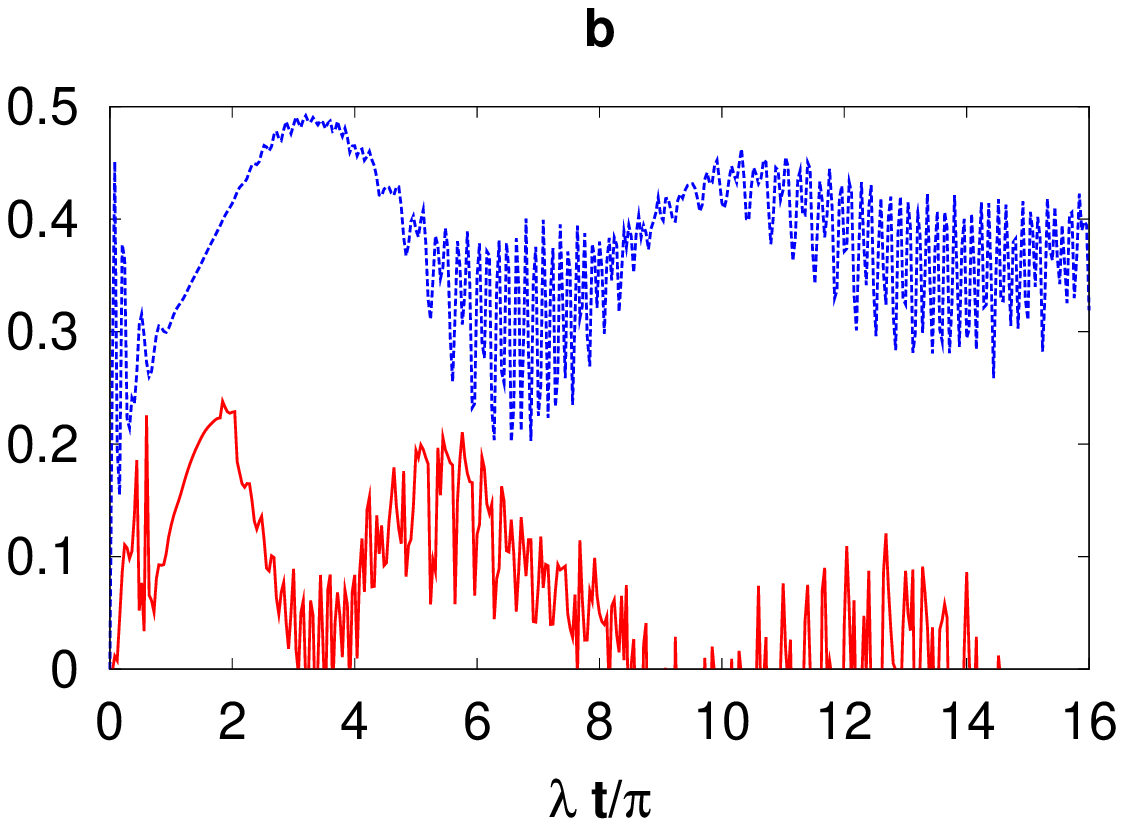}
\includegraphics[width=.45\linewidth]
{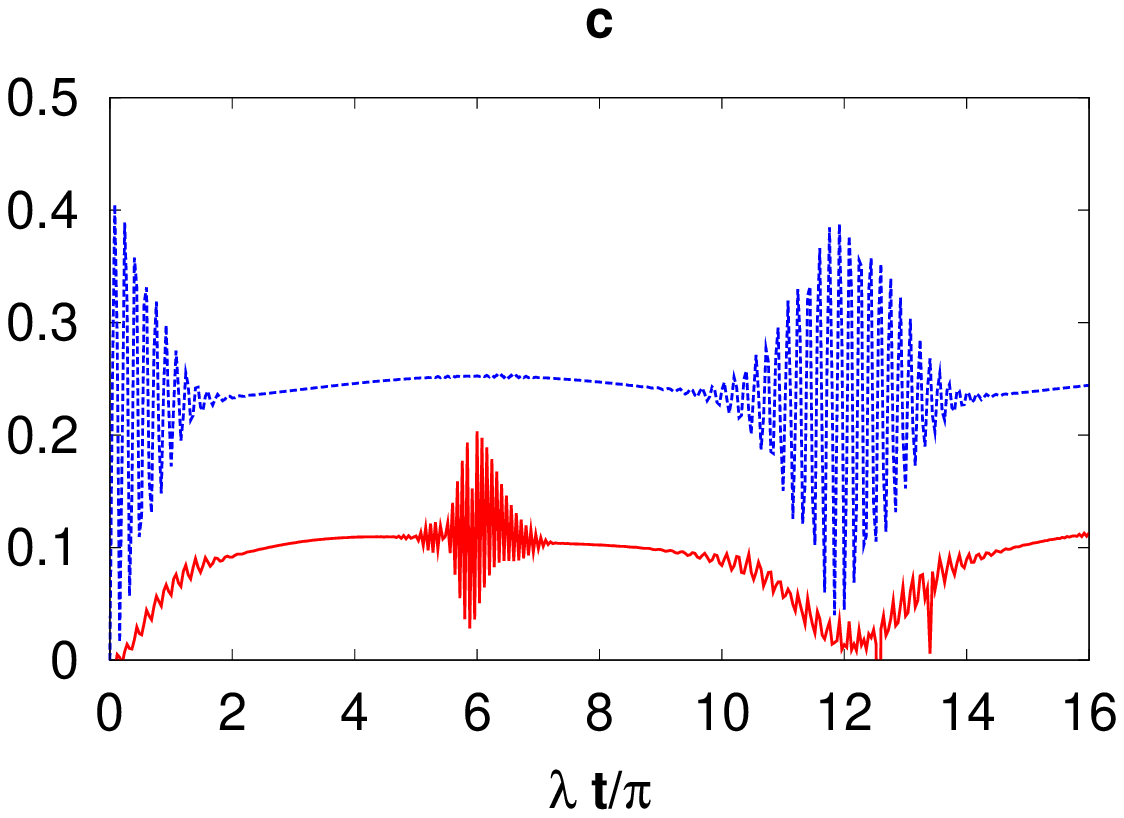}
\includegraphics[width=.45\linewidth]
{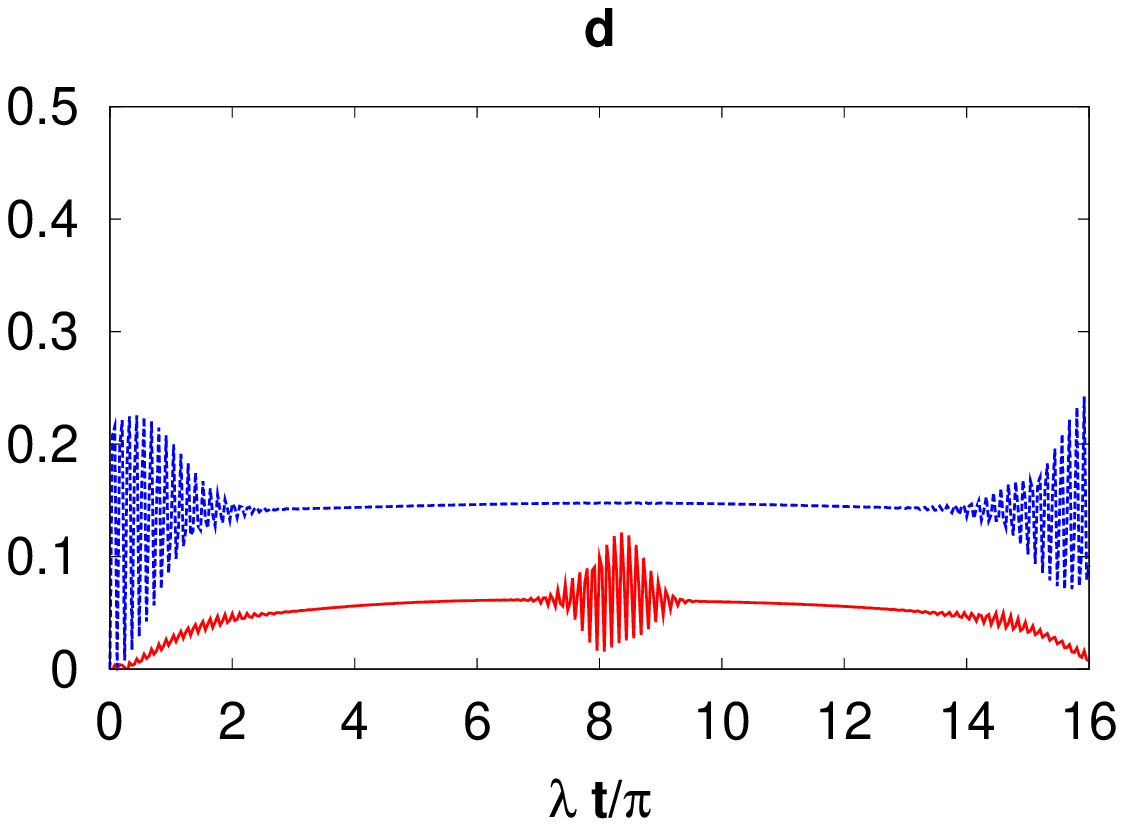}
\includegraphics[width=.45\linewidth]
{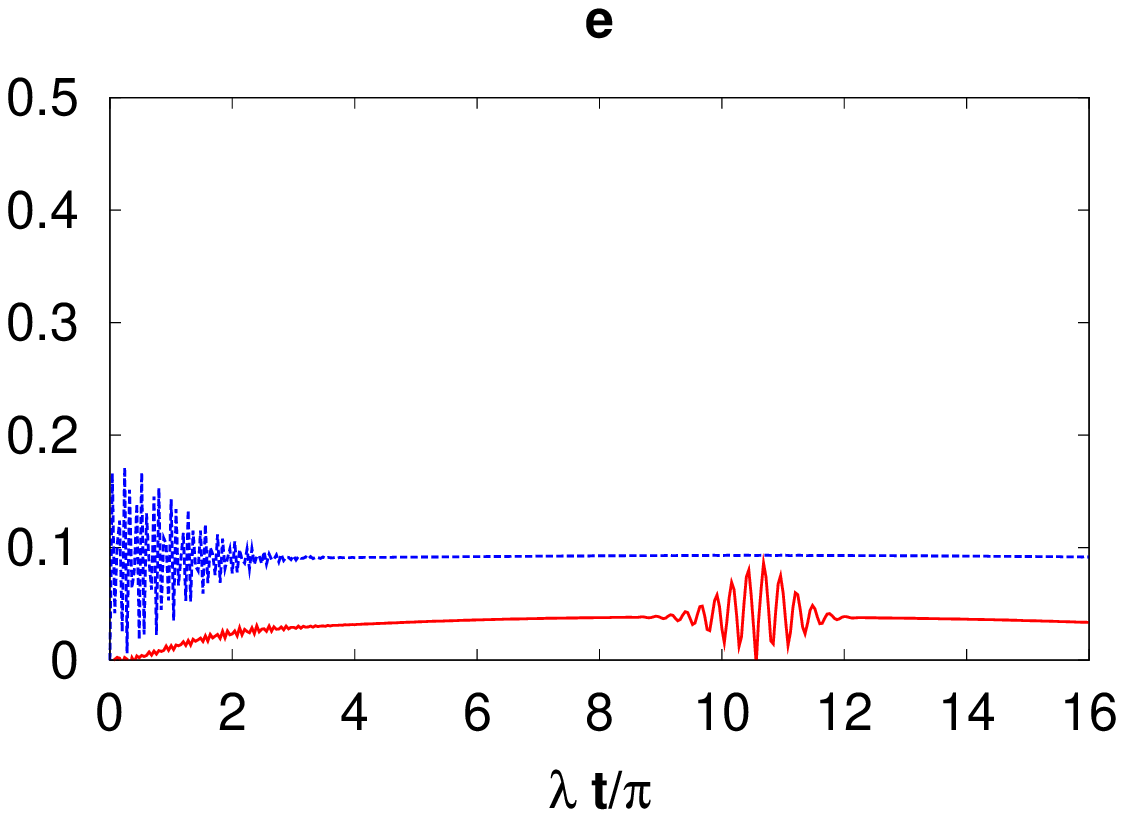}
\end{center}
\caption{The same as Fig. 1 but for $\chi/\lambda=0.0$ where (a) $\delta=0.5$ (b) $\delta=1.0$ (c) $\delta=10.0$ (d) $\delta=15.0$ (e) $\delta=20$  }
\end{figure}
\begin{figure}[tpbh]
\noindent
\begin{center}
\includegraphics[width=.45\linewidth]
{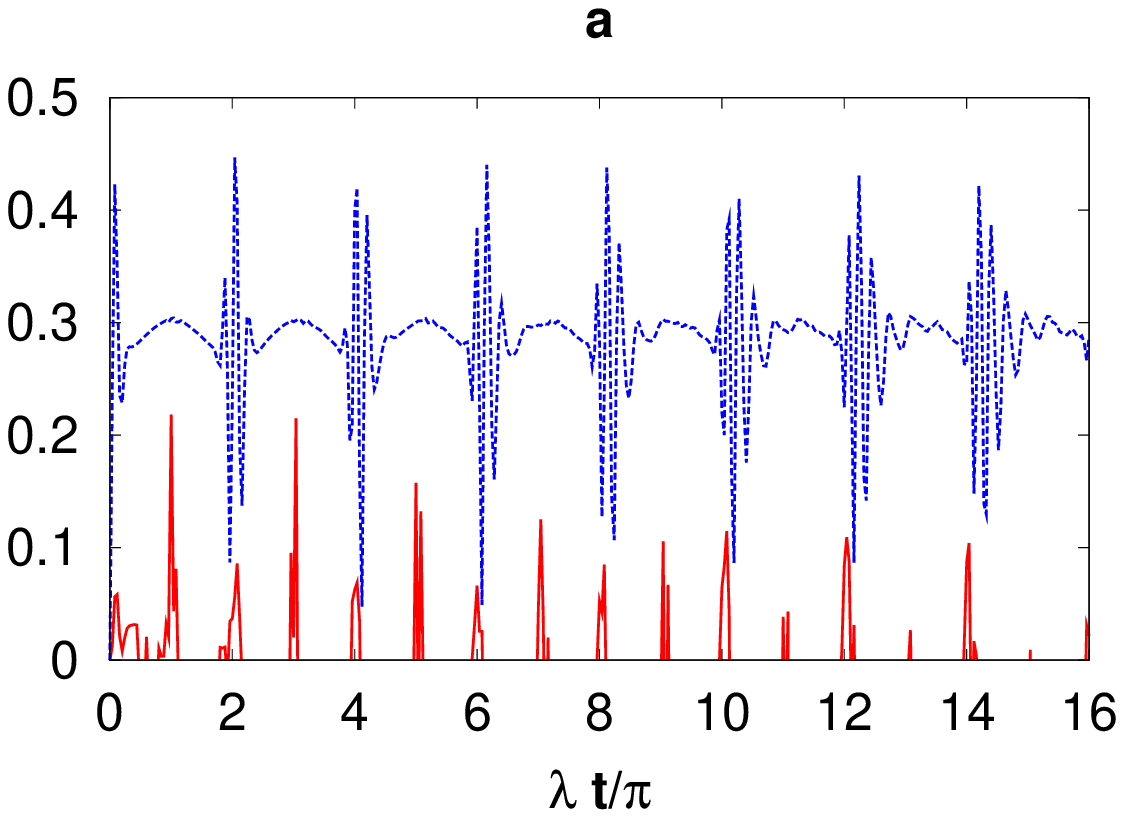}
\includegraphics[width=.45\linewidth]
{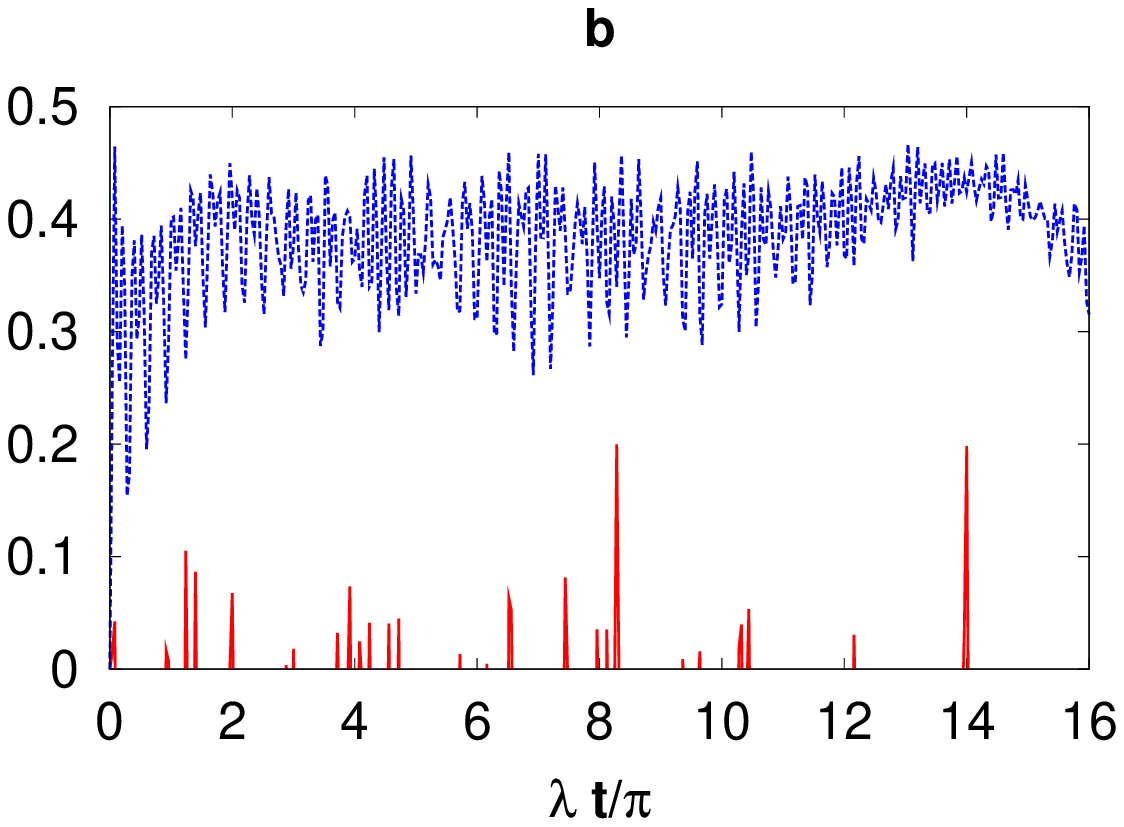}
\end{center}
\caption{The same as Fig. 2 but when $\chi=0.5$ and (a) $\delta=1.0$ (b) $\delta=10.0$  }
\end{figure}
%
%
%
%
%
%
In order to discuss the dynamical feasibility of our scheme in the micromaser regime, in which the lifetime of the cavity is longer than atom field interaction time,  the concurrence $C(\lambda t/\pi)$ and the corresponding total population $\rho_{22}(\lambda t/\pi)+\rho_{33}(\lambda t/\pi)$, as a reference state, are depicted simultaneously in figures 1, 2 and 3 for the Rabi angle $\lambda t/\pi \in [0,16]$.  The mean photon number $\bar n$ is taken about $\bar n=10$, the detuning parameter $\delta=\Delta/\lambda$ carries various values such as $\delta=0.0, 0.5, 1, 10, 15$ and $20$, while the Kerr parameter $\chi/\lambda$ varies such as $\chi/\lambda=0.0, 0.2, 0.5$ and $\chi/\lambda=1.0$. From Fig. 1a, where the interaction occurs resonantly in the absence of the nonlinear media, one can see clearly that the concurrence $C$ exhibits irregular collections of oscillations where the total population oscillates but with amplitudes maximize (no longer exceed the value $0.25$) when $\rho_{22}+\rho_{33}\approx 0.4$; the atoms in the mixed state $\bigl|\begin{array}{l}
\vspace{-2ex}+\\
             -\\
\end{array}\bigr>+\bigl|\begin{array}{l}
\vspace{-2ex}-\\
             +\\
\end{array}\bigr>$. Thus, the lack of the entanglement between the two atoms can be attributed to a large population of the product states $\bigl|\begin{array}{l}
\vspace{-2ex}+\\
             -\\
\end{array}\bigr>$ and  $\bigl|\begin{array}{l}
\vspace{-2ex}-\\
             +\\
\end{array}\bigr>$ which, in turn, decay on the same time scale as $C$. For significantly high average photon number $\bar n$, it was shown that~\cite{BGASMNN07}, quantum effects which are predominant primarily when the photon number is low, help to increase the peak value of the concurrence $C$.
\par On resonance interaction between the atoms and a cavity filled with a Kerr medium of a parameter of order $10^{-1}\lambda$, such as $\chi=0.2 \lambda$, the total population shows pseudo periodical behavior with amplitudes minima decrease as time evolves where the revival period, $t_{R}$, elongates as time goes on. The concurrence $C$ loses its irregular collections where instant oscillations appear after wide time intervals accompany with amplitudes decrease ($\approx 0.3$ in the beginning) as evolves , Fig. 1b. A significant and  interesting distinction in the case when $\chi=0.5\lambda$ is that both $C$ and $\rho_{22}+\rho_{33}$ display a perfect $\pi$ periodicity of oscillations. The total population $\rho_{22}+\rho_{33}$ exhibits collapses and revivals with the revival period $t_{R}$ is shorter and each revival is a perfect carbon copy of its predecessor. The concurrence $C$ shows oscillation packets composed of three peaks with the strongest,$=2.5$, lies at the center where the total population collapses. Moreover, the concurrence $C$ minimizes where the total populations $\rho_{22}+\rho_{33}$ revives while the strong entanglement and collapse of population occur in parallel manner, see Fig. 1c. This periodicity is due to the fact that analytical expressions of both $C$ and $\rho_{22}+\rho_{33}$ can be summed exactly. Thus, the matching value of the Kerr parameter plays more efficient rule than that of quantum effects that leads to increase the peak value of the concurrence $C$~\cite{BGASMNN07}. Opposite results are obtained when a considerable strength of the Kerr medium is applied to the cavity field. In this case, a pseudo periodicity for $C$ is observed where the total population preserves its periodicity accompanied with clear decreasing in its amplitudes. In addition, one can't build a clear insight about the relation between curves of both $C$ and $\rho_{22}+\rho_{33}$, Fig. 1d. Thus, from the above it is found that the Kerr medium can be used to adjust the entanglement degree between qubits, which may provide some reference and even basis for the quantum control of multiple qubits.
\par A surprise is that the case of a non-resonance interaction does not destroy the entanglement phenomenon, see Fig. 2. For small detuning such that $\delta=0.5$, there is no significant change, Fig. 2a, but when $\delta$ increases to unity, the correlations time between the atoms becomes longer with the gradual appearance of nonzero values of $C$, see Fig. 2b. Strictly speaking, the destruction of entanglement occurs for the case when the detuning parameter $\delta$ is absent but not the Kerr parameter $\chi/\lambda$, Fig. 1d.
\par The most intersting and surprising is the case when the inter-atomic transition is considered to be significantly high compared with the cavity frequency such that $\delta=10.0, 15.0$ and $20.0$ where the nonlinear media is considered to be absent. This is revealed in Figs. 2c-e. We see that the total population exhibits the known collape and revival phenomena periodically. Also, we notice that for the increase of detuning parameter $\delta$, the collapses become longer and the beginning strong overlap of the revivals becomes weaker; while the amplitude of the revivals decreases as values of $\delta$ increase. The entanglement evolution divided into three regions. The first one is at the half of the collapse time, where the entanglement degree oscillates with bigger amplitude for short time scale. The oscillation maximum is related crucially to the value of $\delta$, where it occurs at time $\lambda t=(\frac{\delta}{2}+1)\pi$. The last two regions are in the beginning and end of the collape time. In this case, we see that the atom-atom entanglement remain close to its maximum and show intrinsic oscillations only. This is in fact is true for all values of the detuning parameter. The detuning between the field and the atoms leads to an emergence of collapses and revivals, this statement is always true only in the absence of the nonlinear media, see Figs. 2c-e. Overall, we find that the detuning parameter acts as a control parameter for the atom-atom entanglement, where a perfect periodicity can be obtained for suitable high detuning. The revival of the entanglement at time $\lambda t=(\frac{\delta}{2}+1)\pi$ is more easily understood by reference to the density matrix of the system.
As expected, there is no entanglement before the populations of the state $\bigl|\begin{array}{l}
\vspace{-2ex}+\\
             -\\
\end{array}\bigr>+\bigl|\begin{array}{l}
\vspace{-2ex}-\\
             +\\
\end{array}\bigr>$
and the state $\bigl|\begin{array}{l}
\vspace{-2ex}+\\
             +\\
\end{array}\bigr>+\bigl|\begin{array}{l}
\vspace{-2ex}-\\
             -\\
\end{array}\bigr>$ depopulates, but some entanglement appears for longer times, Figs. 2b-e, and the concurrence becomes equal to the population of the state $\bigl|\begin{array}{l}
\vspace{-2ex}+\\
             -\\
\end{array}\bigr>+\bigl|\begin{array}{l}
\vspace{-2ex}-\\
             +\\
\end{array}\bigr>$, see Fig. 1c. \\
\section{injection of a ground atom after the excited one}
\label{sec:5}
In this case, the time dependent state vector of the full system can be expressed in the form
\begin{equation}
\label{eq68}
|\psi_{AAF}(t)\rangle=\biggl| S_{\begin{array}{l}
\vspace{-2ex}+\\
             -\\
\end{array}
}(t)\biggr >\mid n\rangle+\biggl| S_{\begin{array}{l}
\vspace{-2ex}+\\
             +\\
\end{array}
}(t)\biggr >\mid n-1\rangle+\biggl| S_{\begin{array}{l}
\vspace{-2ex}-\\
             -\\
\end{array}
}(t)\biggr >\mid n+1\rangle+\biggl| S_{\begin{array}{l}
\vspace{-2ex}-\\
             +\\
\end{array}
}(t)\biggr >\mid n\rangle,
\end{equation}
where
\begin{equation}
\label{eq69}
\biggl| S_{\begin{array}{l}
\vspace{-2ex}+\\
             -\\
\end{array}
}(t)\biggr >=\sum_{n} C^{n}~\Gamma_{1}(n,t)\Gamma_{1}^{\ast}(n-1,t)\bigl|\begin{array}{l}
\vspace{-2ex}+\\
             -\\
\end{array}\bigr>,
\end{equation}
\begin{equation}
\label{eq70}
\biggl| S_{\begin{array}{l}
\vspace{-2ex}+\\
             +\\
\end{array}
}(t)\biggr >=\sum_{n} C^{n}~\Gamma_{1}(n,t)\Gamma_{2}(n,t)\bigl|\begin{array}{l}
\vspace{-2ex}+\\
             +\\
\end{array}\bigr>,
\end{equation}
\begin{equation}
\label{eq71}
\biggl| S_{\begin{array}{l}
\vspace{-2ex}-\\
             -\\
\end{array}
}(t)\biggr >=\sum_{n} C^{n}~\Gamma_{1}^{\ast}(n+1,t)\Gamma_{2}(n+1,t)\bigl|\begin{array}{l}
\vspace{-2ex}-\\
             -\\
\end{array}\bigr>,
\end{equation}
\begin{equation}
\label{eq72}
\biggl| S_{\begin{array}{l}
\vspace{-2ex}-\\
             +\\
\end{array}
}(t)\biggr >=\sum_{n} C^{n}~[\Gamma_{2}(n+1,t)]^{2}\bigl|\begin{array}{l}
\vspace{-2ex}-\\
             +\\
\end{array}\bigr>,
\end{equation}
Following the same procedure as in the previous section, and after using the notations
$\bigl|\begin{array}{l}
\vspace{-2ex}+\\
             -\\
\end{array}\bigr>\equiv\mid 1\rangle$, $\bigl|\begin{array}{l}
\vspace{-2ex}+\\
             +\\
\end{array}\bigr>\equiv\mid 2\rangle$, $\bigl|\begin{array}{l}
\vspace{-2ex}-\\
             -\\
\end{array}\bigr>\equiv\mid 3\rangle$, $\bigl|\begin{array}{l}
\vspace{-2ex}-\\
             +\\
\end{array}\bigr>\equiv\mid 4\rangle$, the elements of the reduced atomic density matrix (\ref{eq57}) read
\begin{equation}
\label{eq73}
\rho_{11}(t)=\sum_{n}|C^{n}|^{2}|\Gamma_{1}(n,t)|^{2}|\Gamma_{1}(n-1,t)|^{2},
\end{equation}
\begin{equation}
\label{eq74}
\rho_{12}(t)=\sum_{n}C^{n}C^{\ast n+1}\Gamma_{1}(n,t)\Gamma_{1}^{\ast}(n-1,t)\Gamma_{1}^{\ast}(n+1,t)\Gamma_{2}^{\ast}(n+1,t),
\end{equation}
\begin{equation}
\label{eq75}
\rho_{13}(t)=\sum_{n}C^{n+k}C^{\ast n}|\Gamma_{1}(n,t)|^{2}\Gamma_{1}(n+1,t) \Gamma_{2}^{\ast}(n+1,t),
\end{equation}
\begin{equation}
\label{eq76}
\rho_{14}(t)=\sum_{n}|C^{n}|^{2} \Gamma_{1}(n,t)\Gamma_{1}^{\ast}(n-1,t)[\Gamma_{2}^{\ast}(n+1,t)]^{2},
\end{equation}
\begin{equation}
\label{eq77}
\rho_{22}(t)=\sum_{n}|C^{n}|^{2}|\Gamma_{1}(n,t)|^{2}|\Gamma_{2}(n,t)|^{2},
\end{equation}
\begin{equation}
\label{eq78}
\rho_{23}(t)=\sum_{n}C^{n+2}C^{\ast n} \Gamma_{1}(n+2,t)\Gamma_{2}(n+2,t)\Gamma_{1}(n,t) \Gamma_{2}^{\ast}(n+1,t),
\end{equation}
\begin{equation}
\label{eq79}
\rho_{24}(t)=\sum_{n}C^{n+2}C^{\ast n} |\Gamma_{2}(n+1,t)|^{2}\Gamma_{1}(n+1,t)\Gamma_{2}^{\ast}(n+1,t),
\end{equation}
\begin{equation}
\label{eq80}
\rho_{33}(t)=\sum_{n}|C^{n}|^{2}|\Gamma_{1}(n,t)|^{2}|\Gamma_{2}(n+1,t)|^{2},
\end{equation}
\begin{equation}
\label{eq81}
\rho_{34}(t)=\sum_{n}C^{n}C^{\ast n+1} \Gamma_{1}^{\ast}(n,t) \Gamma_{2}(n+1,t)[\Gamma_{2}^{\ast}(n+2,t)]^{2},
\end{equation}
\begin{equation}
\label{eq82}
\rho_{44}(t)=\sum_{n}|C^{n}|^{2}|\Gamma_{2}(n+1,t)|^{4},
\end{equation}
\begin{figure}[tpbh]
\noindent
\begin{center}
\includegraphics[width=.45\linewidth]
{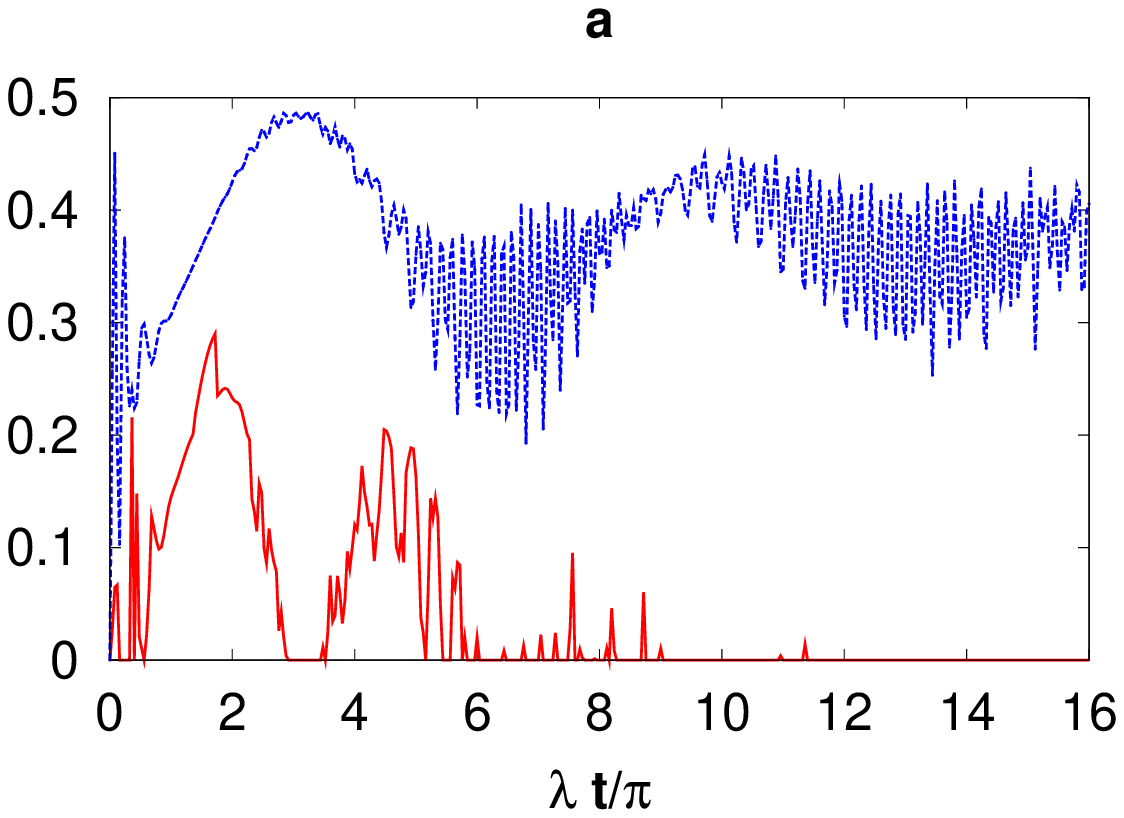}
\includegraphics[width=.45\linewidth]
{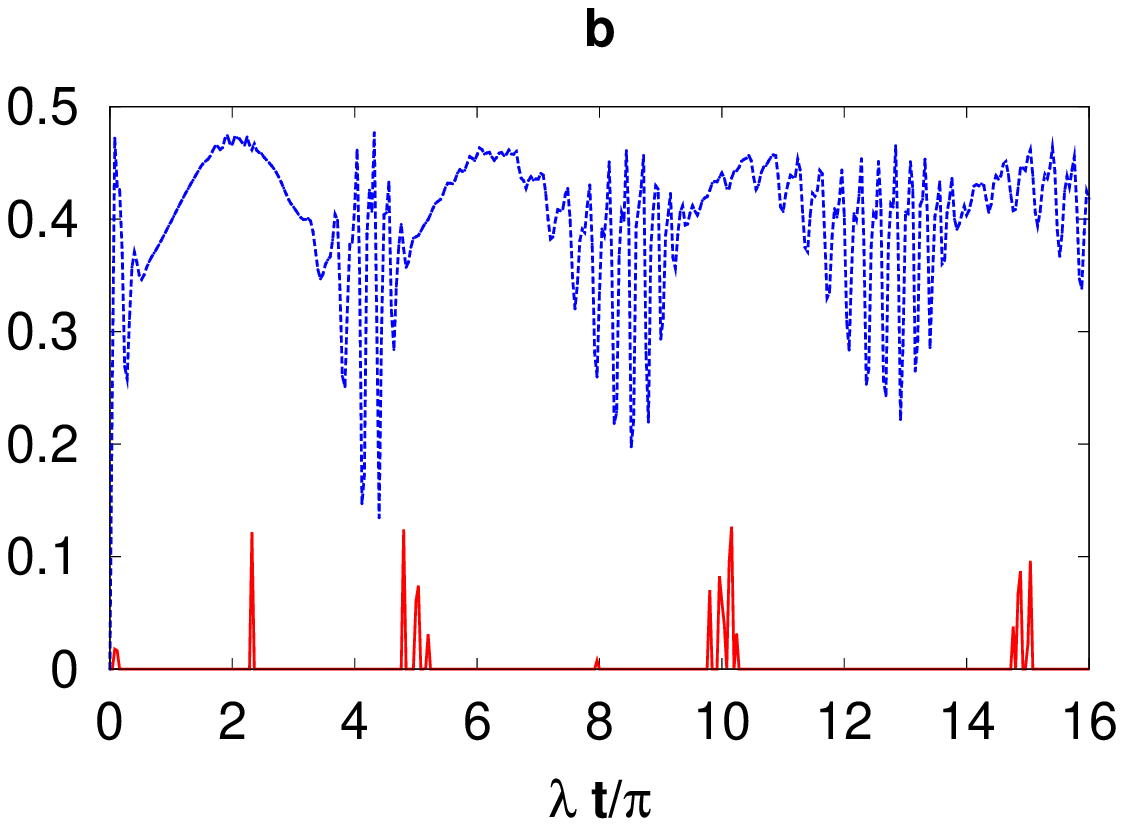}
\includegraphics[width=.45\linewidth]
{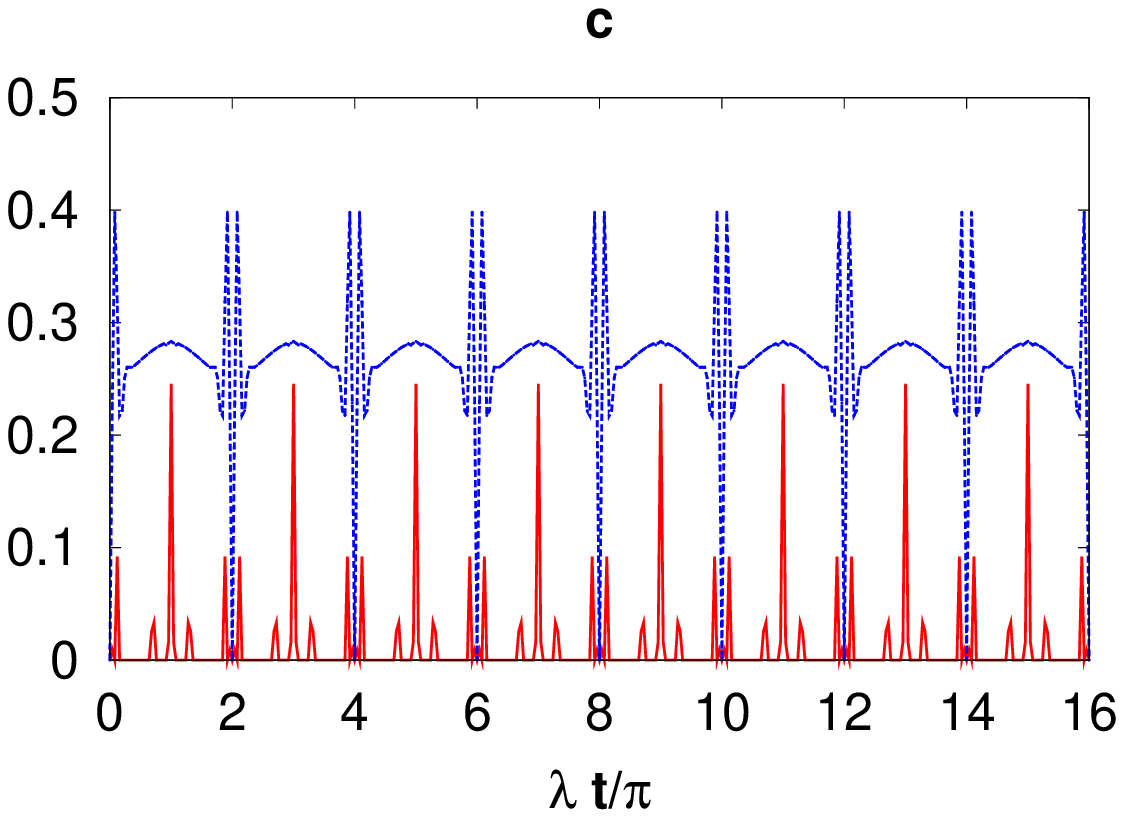}
\includegraphics[width=.45\linewidth]
{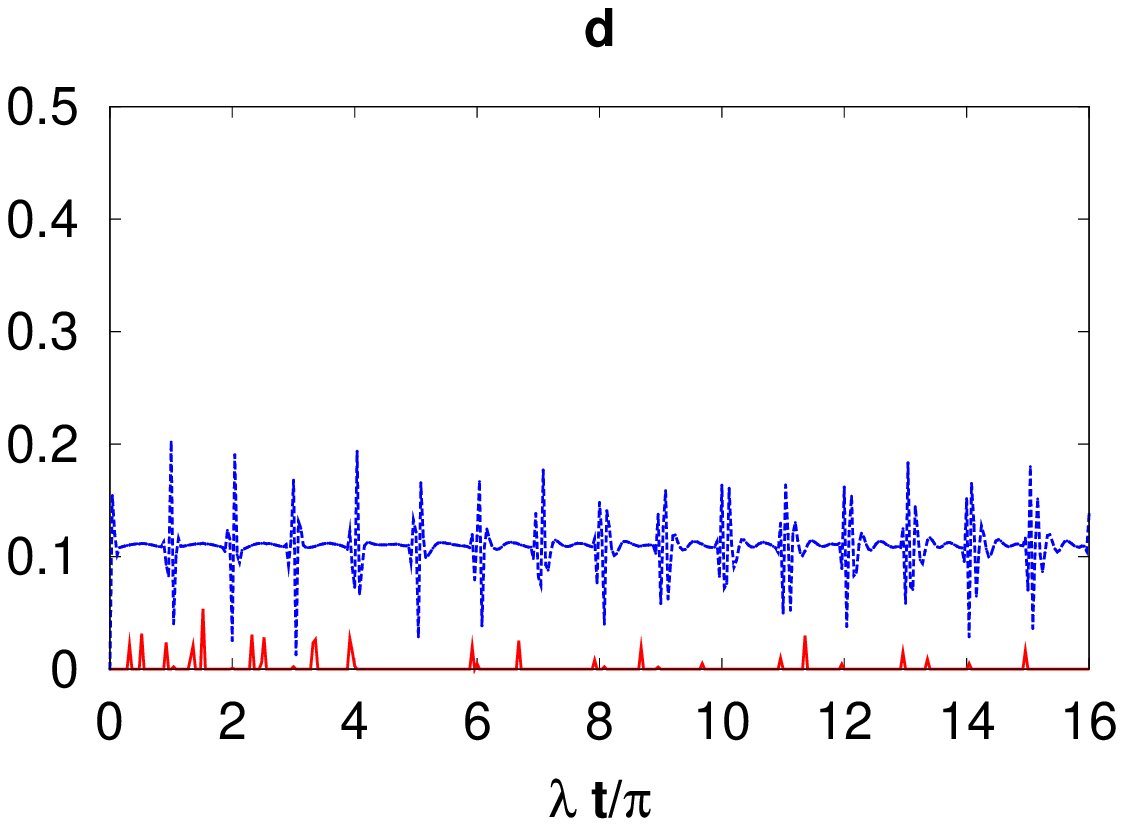}
\end{center}
\caption{ The concurrence $C(\lambda t/\pi)$ (solid red curve) and the total population $\rho_{22}(\lambda t/\pi)+\rho_{33}(\lambda t/\pi)$ (dashed blue curve) for $\bar{n}=10$, $\delta=0.0$ where (a) $\chi/\lambda=0.0$ (b) $\chi/\lambda=0.2$ (c) $\chi/\lambda=0.5$ (c) $\chi/\lambda=1.0$ }
\end{figure}
\begin{figure}[tpbh]
\noindent
\begin{center}
\includegraphics[width=.45\linewidth]
{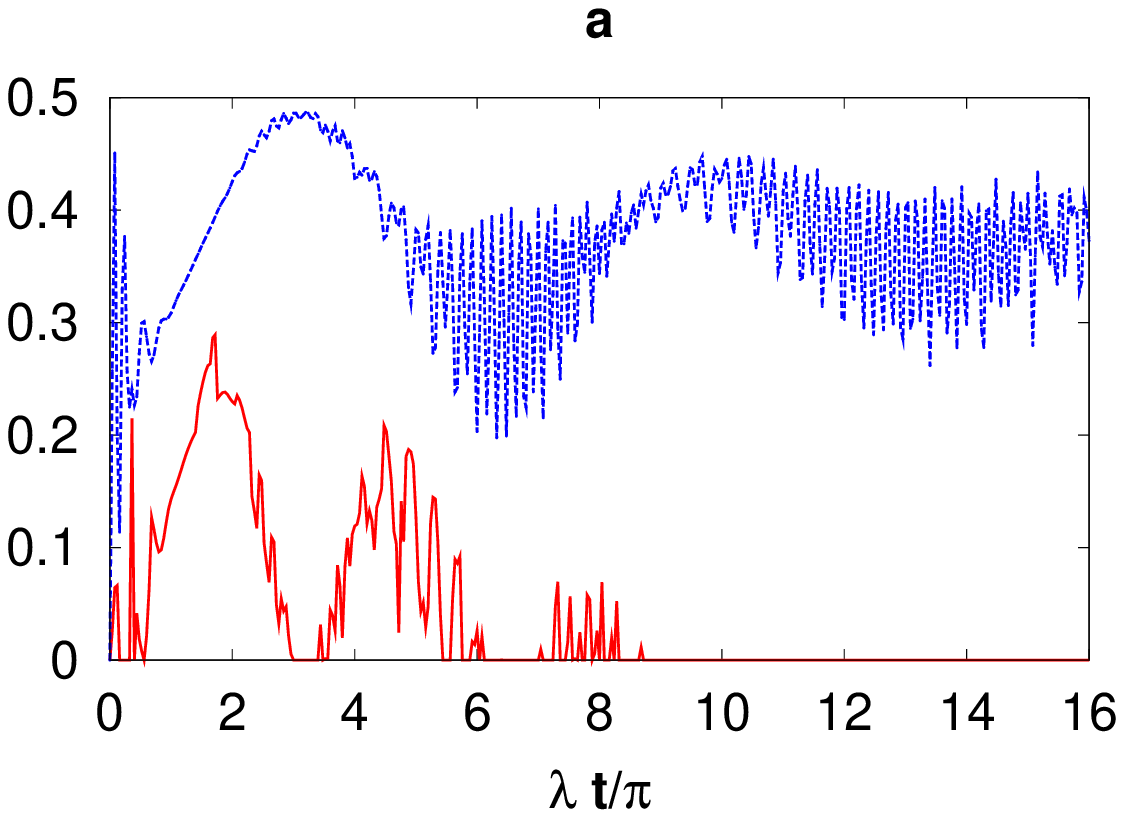}
\includegraphics[width=.45\linewidth]
{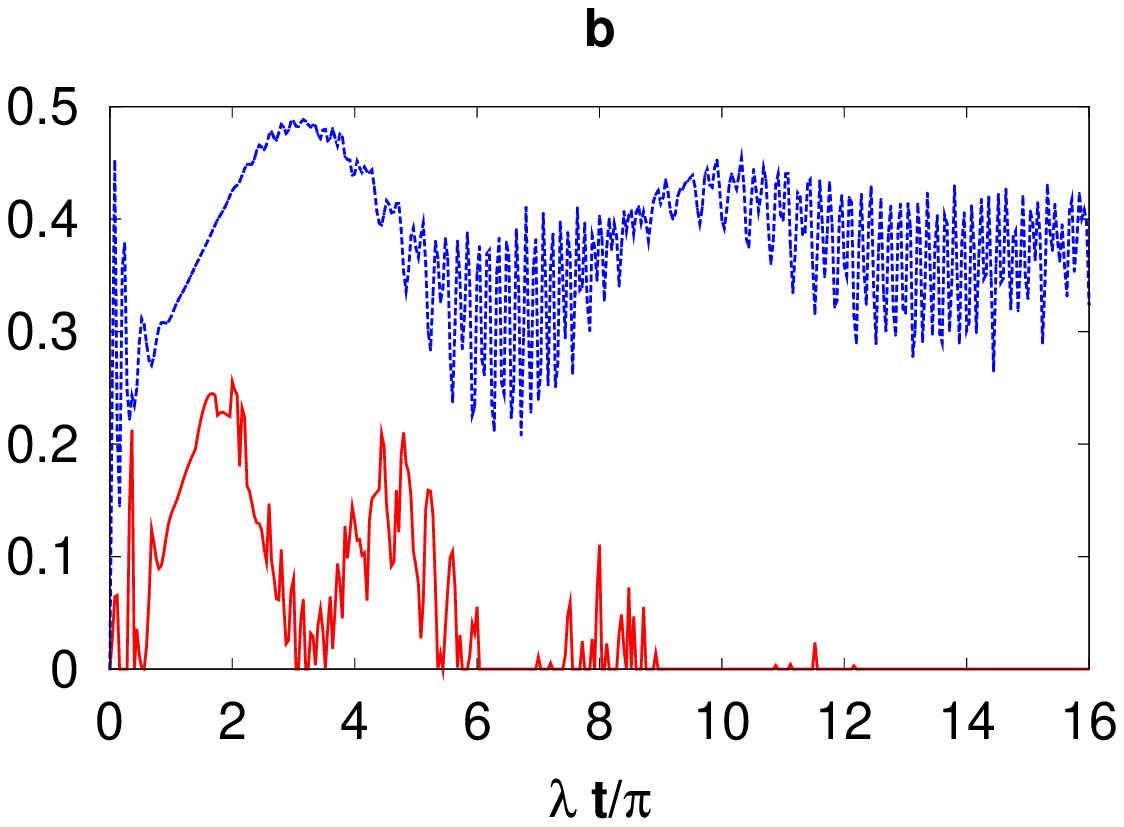}
\includegraphics[width=.45\linewidth]
{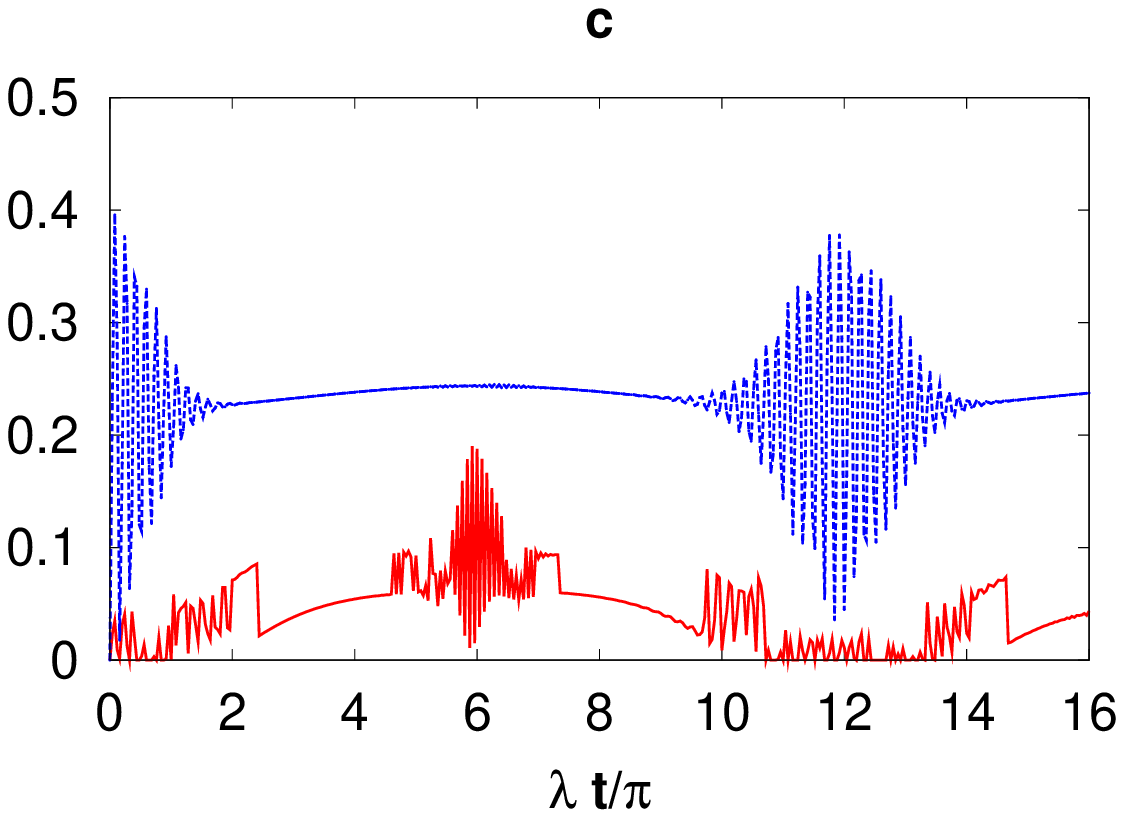}
\includegraphics[width=.45\linewidth]
{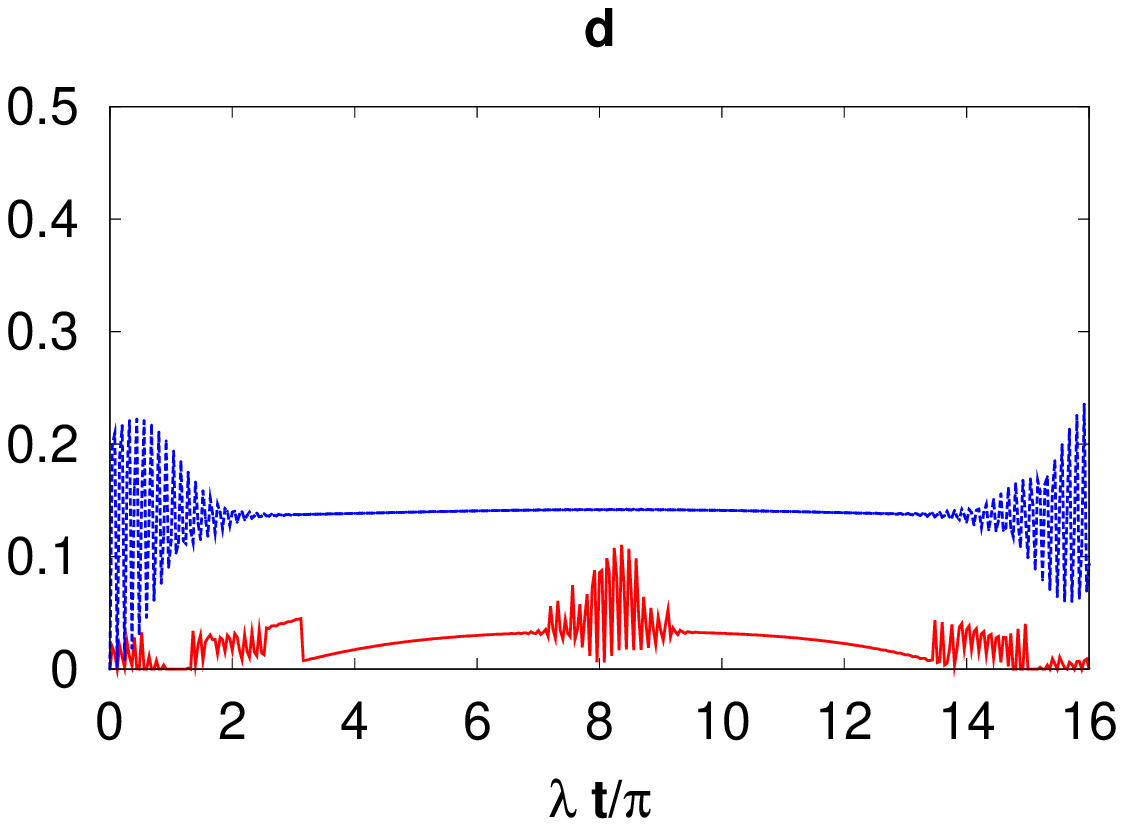}
\includegraphics[width=.45\linewidth]
{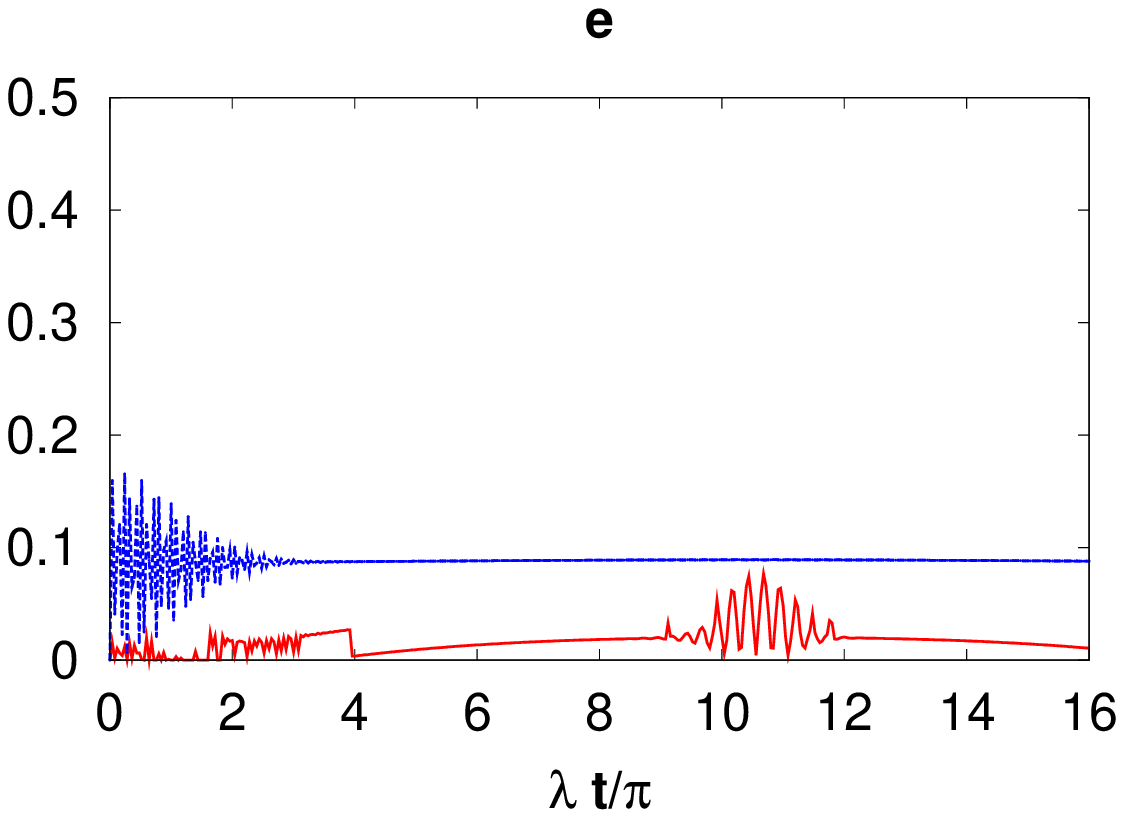}
\end{center}
\caption{The same as Fig. 4 but for $\chi/\lambda=0.0$ where (a) $\delta=0.5$ (b) $\delta=1.0$ (c) $\delta=10.0$ (d) $\delta=15.0$ (e) $\delta=20$}
\end{figure}
\begin{figure}[tpbh]
\noindent
\begin{center}
\includegraphics[width=.45\linewidth]
{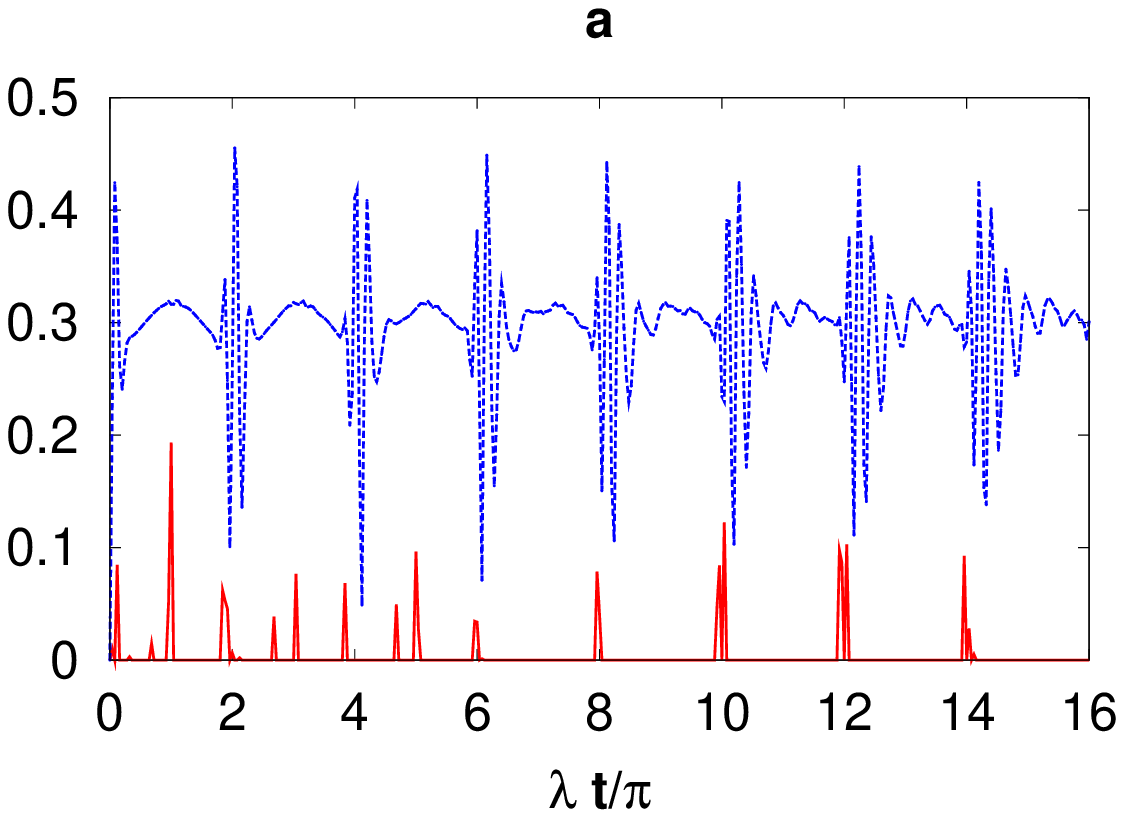}
\includegraphics[width=.45\linewidth]
{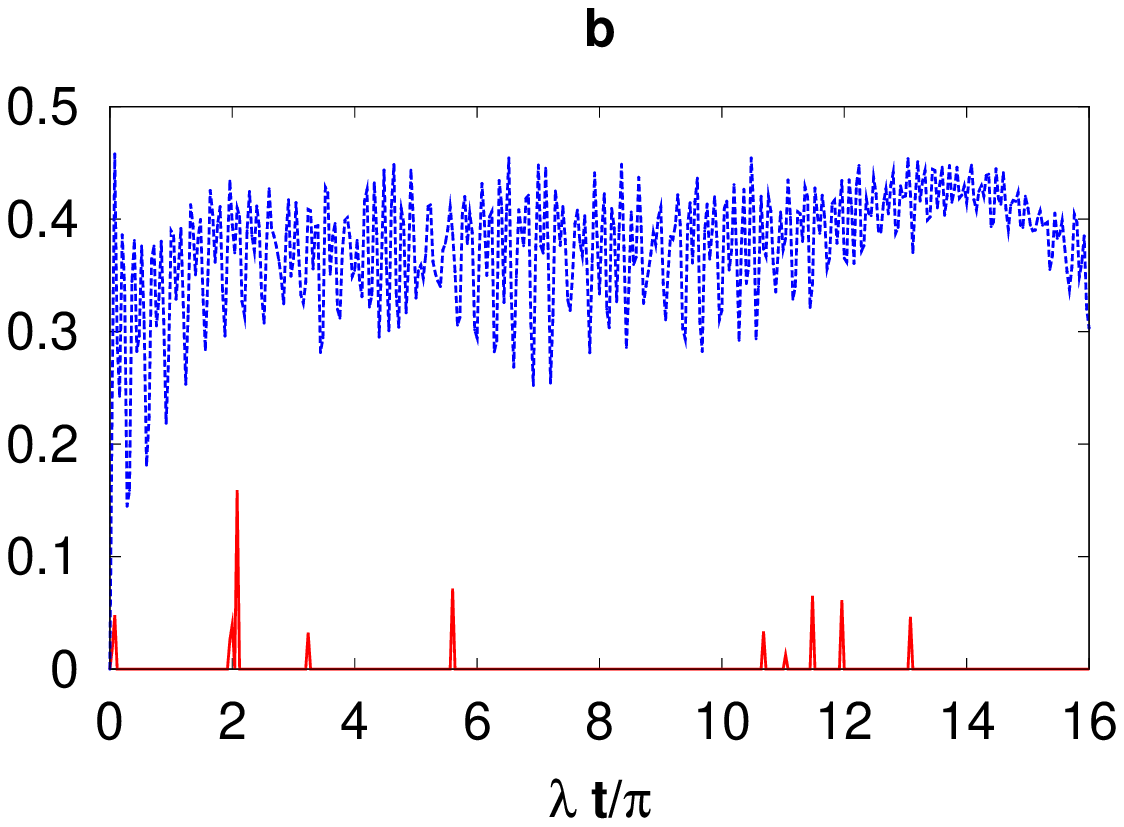}
\end{center}
\caption{The same as Fig. 5 but when $\chi=0.5$ and (a) $\delta=1.0$ (b) $\delta=10.0$}
\end{figure}
To better understand the situation and gain clearer insight into the entanglement dynamics, hereafter we shall analyze the evolution of the concurrence $C(\lambda t/\pi)$ and the total populations $\rho_{22}(\lambda t/\pi)+\rho_{33}(\lambda t/\pi)$ where the parametres remain as the same in the past section. 
\par Generally, a similar conclusion is reached for the present case but with considerable distinctions in its details. As in the case of two excited atoms, there is no initial entanglement between the
atoms, and at early times the entanglement builds up rapidly to a maximum
appearing at short time $\lambda t\approx 2.5\pi$, Fig. 4a. However, comparing with the case of two successive excited atoms, shown in the previous section, we see that the entanglement maxima that obtained in the present case are generally lower than that obtained with two successive excited atoms, see Figs. 1-3 and 4-6. It is interesting to note that after passing through the maximum, the entanglement decays with two different time scales, Figs. 2 and 5. This effect is more pronounced if we choose the interaction with atoms to be initially in different quantum states. Thus, atom-atom entanglement depends crucially on the coherent transfer of the population between the atoms states, specially in the presence of a nonlinear media as well as the detuning parameter $\delta$. In this process, the population is efficiently transferred from the more populated state, $\bigl|\begin{array}{l}
\vspace{-2ex}+\\
             +\\
\end{array}\bigr>+\bigl|\begin{array}{l}
\vspace{-2ex}-\\
             -\\
\end{array}\bigr>$ to the state, $\bigl|\begin{array}{l}
\vspace{-2ex}+\\
             -\\
\end{array}\bigr>+\bigl|\begin{array}{l}
\vspace{-2ex}-\\
             +\\
\end{array}\bigr>$ before it decays to the initial state leading to the enhancement of the entanglement.\\
\section{Phase spase distribution}
\label{sec:6}
We devote this section to concentrate on a function, which considered an important one of the functions, $W$, (Husimi) $Q$, and (Glauber-Sudershan) $P$ functions, that is the Husimi $Q$-function which has the nice property of being always positive and further advantage of being readily measurable by quantum tomographic techniques~\cite{BOTAWA98, MANTOM97}. It has been shown from earlier studies~\cite{EWIGNER32, ZWIGNer32, KAGL69, HICOSCWG84} that these quasi-probability functions are important for the statistical description of a microscopic system and provide insight into the nonclassical features of the radiation fields. The homodyne measurement of an electromagnetic field gives all possible linear combinations of the field quadratures. The average of the random outcomes of the measurement is connected with the marginal distribution of any quasi-probability used in quantum optics. In fact, Husimi $Q$-function is not only a convenient tool to calculate the expectation values of anti-normally ordered products of operators, but also to give some insight into the mechanism of interaction for the model under consideration. The relation between the phase-space measurement; Husimi $Q$-function; and the classical information-theoretic entropy associated with quantum fields was introduced by Wehrl~\cite{WEHRL79}, which referred to as the Shannon information of the Husimi Q-function. Thus, Husimi $Q$-function can be related to quantum entanglement in different approaches~\cite{WEHRL79, PEKRPELUSZ86, FAGU06, CAALCARA09, HUFAN09, MIMAWA00, MIWAIM01, BERETA84}. Furthermore, separable state is represented by a localized wave packet in phase space. Since coherent states are the most localized states in the Husimi representation, it is argued that~\cite{SUGITA03} delocalization of the Husimi distribution implies correlation-hence entanglement- between system particles.
\\
The Husimi $Q$-function can be given in the form as
~\cite{HICOSCWG84, HUSIMI40, FuSOLO001}
\begin{equation}
\label{eq83}
Q(\alpha)=\frac{\langle \alpha\mid\rho_{F}\mid\alpha\rangle}{\pi},
\end{equation}
where $\rho_{F}$ is the reduced density operator of the cavity field given by tracing over the atomic variables of the full density operator (\ref{eq17}). The state $\mid\alpha\rangle$ represents the well-known coherent state with amplitude $\alpha=X+i Y$. Inserting the obtained $\rho_{F}$ into Eq. (\ref{eq83}), we
can easily obtain the Husimi $Q$-function of the cavity field

\begin{equation}
\label{eq84}
Q=\frac{1}{\pi}(\langle\alpha,\begin{array}{l}
\vspace{-2ex}+\\
             +\\
\end{array}\mid\rho_{F}\mid\alpha,\begin{array}{l}
\vspace{-2ex}+\\
             +\\
\end{array}\rangle+\langle\alpha,\begin{array}{l}
\vspace{-2ex}-\\
             -\\
\end{array}\mid\rho_{F}\mid\alpha,\begin{array}{l}
\vspace{-2ex}-\\
             -\\
\end{array}\rangle),
\end{equation}
where 
\begin{equation}
\label{eq85}
\mid\alpha\rangle=e^{-\mid\alpha_{0}\mid^{2}/2}\sum_{n=0}^{\infty}\frac{\alpha_{0}^{n}}{\sqrt{n!}},
\end{equation}
\subsection{Injection of two excited atoms one by one}
\label{subsec:6.1}
In this case, the $Q$-function is given by
\begin{equation}
\label{eq86}
Q=\frac{1}{\pi}\biggl(\biggl|\biggl<\alpha,\begin{array}{l}
\vspace{-2ex}+\\
             +\\
\end{array}\biggl|U_{\begin{array}{l}
\vspace{-2ex}+\\
             +\\
\end{array}},n\biggr>\biggr|^{2}+\biggl|\biggl<\alpha,\begin{array}{l}
\vspace{-2ex}-\\
             -\\
\end{array}\biggl|U_{\begin{array}{l}
\vspace{-2ex}-\\
             -\\
\end{array}},n+1\biggr>\biggr|^{2}\biggr),
\end{equation}
where for the states $\mid\begin{array}{l}
\vspace{-2ex}+\\
             +\\
\end{array}\rangle$ and $\mid\begin{array}{l}
\vspace{-2ex}-\\
             -\\
\end{array}\rangle$, $\rho_{F}$ reads
\begin{equation}
\label{eq87}
\rho_{F}=\biggl| U_{\begin{array}{l}
\vspace{-2ex}+\\
             +\\
\end{array}
}(t),n\biggr >\biggl< U_{\begin{array}{l}
\vspace{-2ex}+\\
             +\\
\end{array}
}(t),n\biggr| +\biggl| U_{\begin{array}{l}
\vspace{-2ex}-\\
             -\\
\end{array}
}(t),n+1\biggr >\biggl< U_{\begin{array}{l}
\vspace{-2ex}-\\
             -\\
\end{array}
}(t),n+1\biggr|,
\end{equation}
in this case
\begin{equation}
\label{eq88}
\biggl<\alpha,\begin{array}{l}
\vspace{-2ex}+\\
             +\\
\end{array}\biggl| U_{\begin{array}{l}
\vspace{-2ex}+\\
             +\\
\end{array}
}(t),n\biggr >=e^{-(\mid\alpha\mid^{2}+\mid\alpha_{0}\mid^{2})/2}\sum_{n}\frac{(\alpha_{0}\alpha^{\ast})^{n}}{n!}[\Gamma_{1}(n,t)]^{2},
\end{equation}
\begin{equation}
\label{eq89}
\biggl<\alpha,\begin{array}{l}
\vspace{-2ex}-\\
             -\\
\end{array}\biggl| U_{\begin{array}{l}
\vspace{-2ex}-\\
             -\\
\end{array}
}(t),n+1\biggr >=e^{-(\mid\alpha\mid^{2}+\mid\alpha_{0}\mid^{2})/2}\sum_{n}\frac{(\alpha_{0}\alpha^{\ast})^{n}\alpha^{\ast 2}}{\sqrt{n!(n+2)!}}\Gamma_{2}(n+1,t)\Gamma_{2}(n+2,t),
\end{equation}

\begin{equation}
\label{eq90}
\alpha=X+iY.
\end{equation}
\begin{figure}[tpbh]
\noindent
\begin{center}
\includegraphics[width=.3\linewidth]
{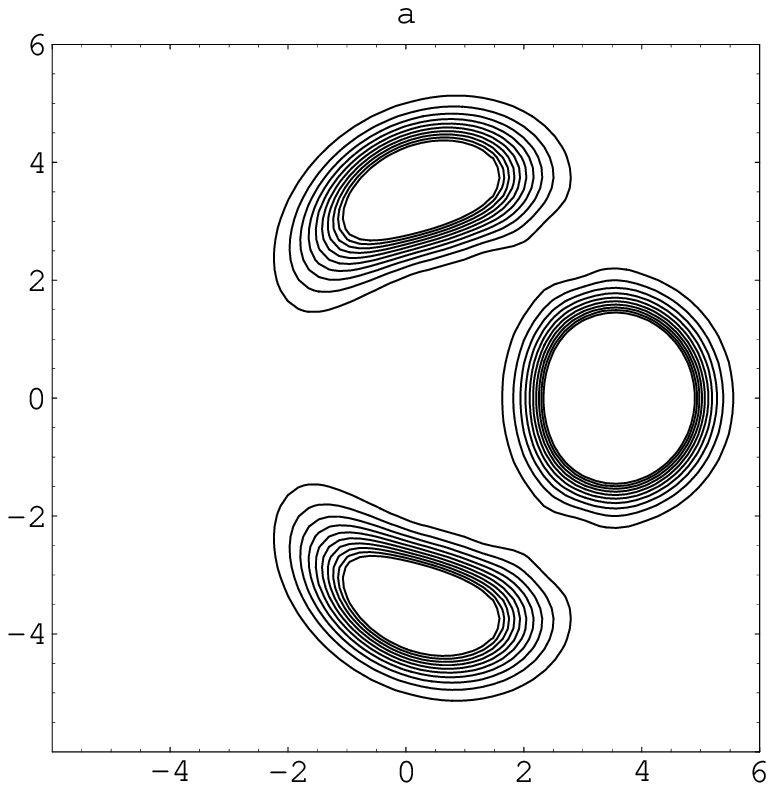}
\includegraphics[width=.3\linewidth]
{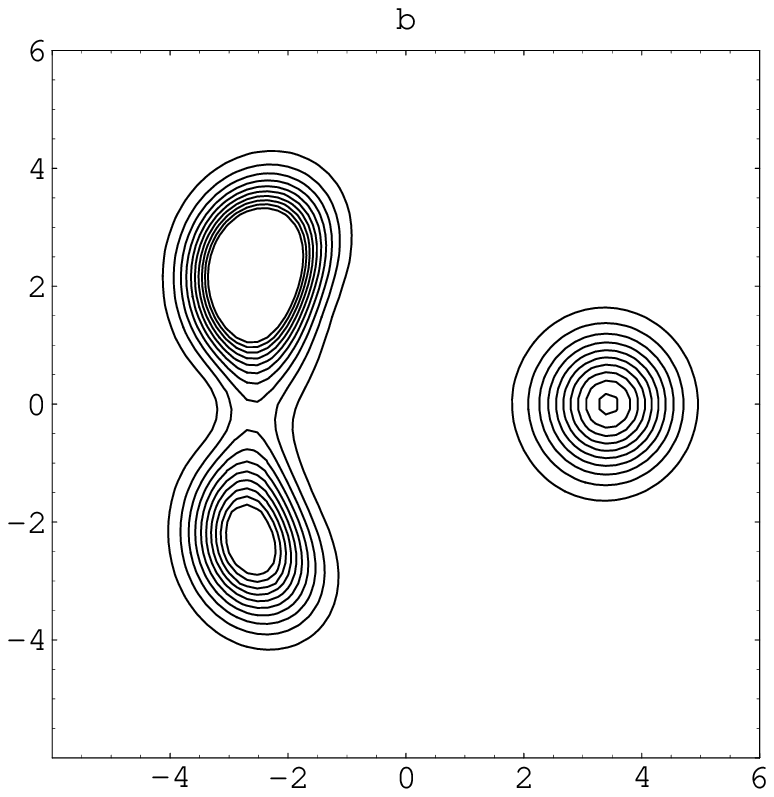}\\
\includegraphics[width=.3\linewidth]
{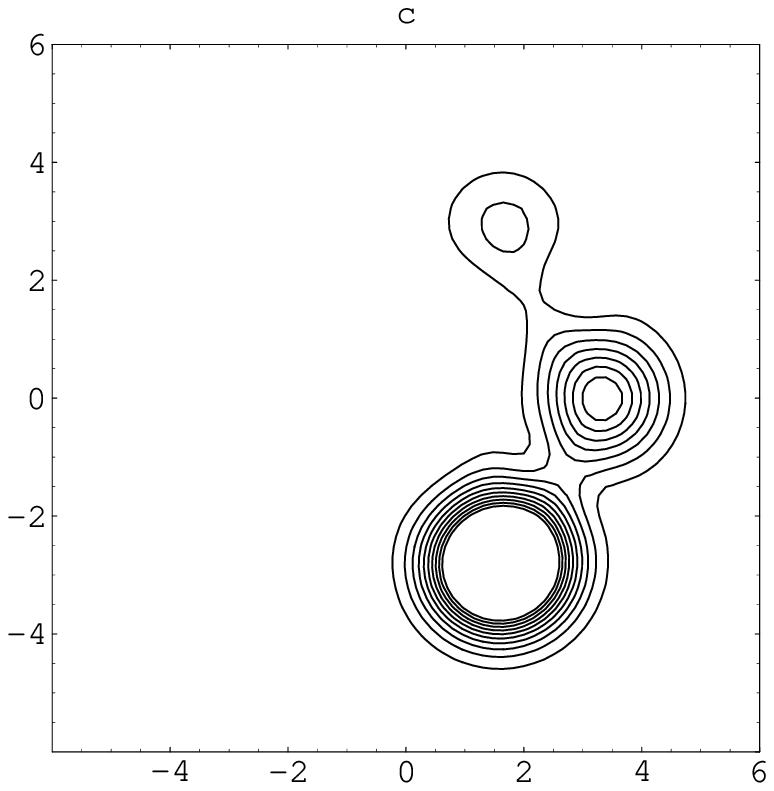}
\includegraphics[width=.3\linewidth]
{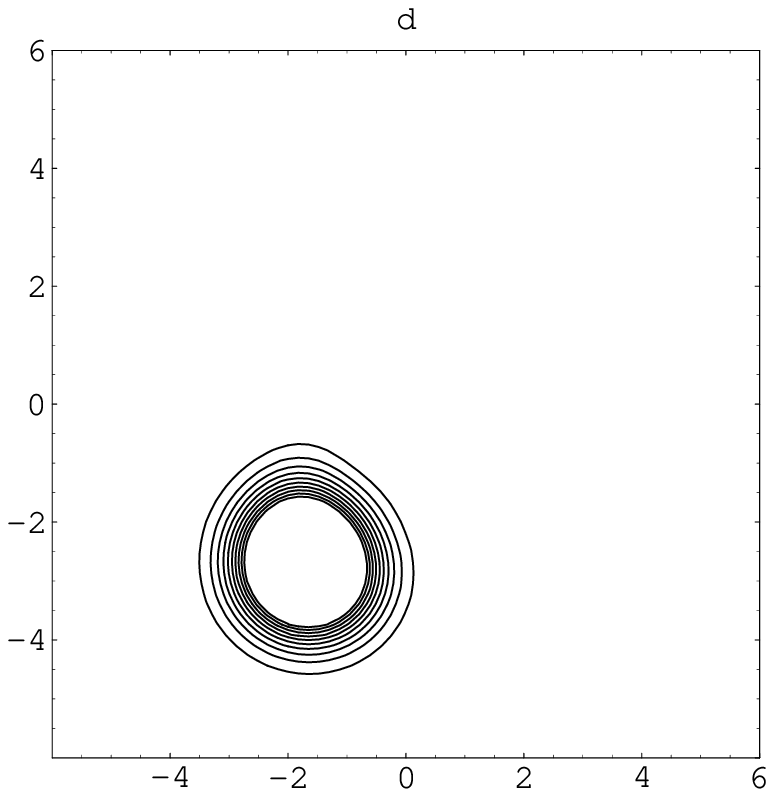}
\end{center}
\caption{Conour plots for Husimi $Q$-function  for $\bar{n}=10$, $\delta=0.0$, $\lambda t=5\pi/3$ where (a) $\chi/\lambda=0.0$ (b) $\chi/\lambda=0.2$ (c) $\chi/\lambda=0.5$ (d) $\chi/\lambda=1.0$  }
\end{figure}
\begin{figure}[tpbh]
\noindent
\begin{center}
\includegraphics[width=.3\linewidth]
{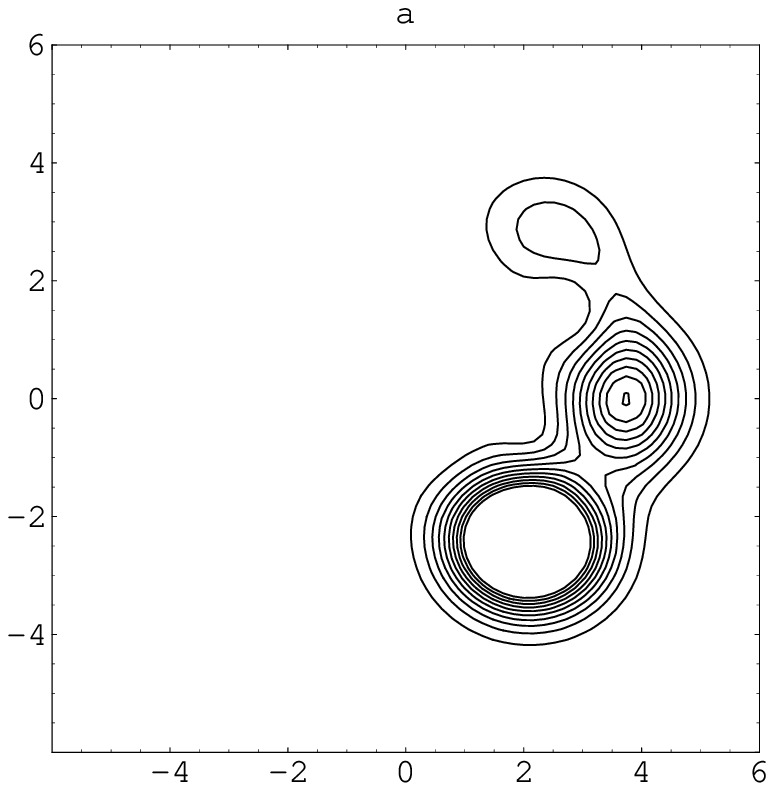}
\includegraphics[width=.3\linewidth]
{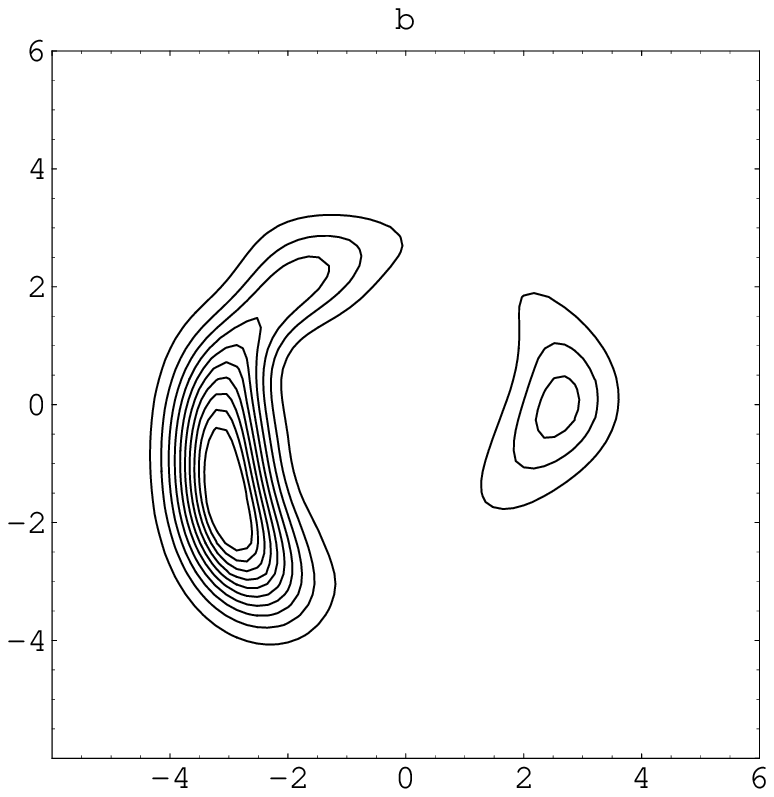}
\end{center}
\caption{The same as Fig. 7 but when $\delta=10.0$ and (a) $\chi/\lambda=0.0$ (b) $\chi/\lambda=1.0$  }
\end{figure}
\begin{figure}[tpbh]
\noindent
\begin{center}
\includegraphics[width=.3\linewidth]
{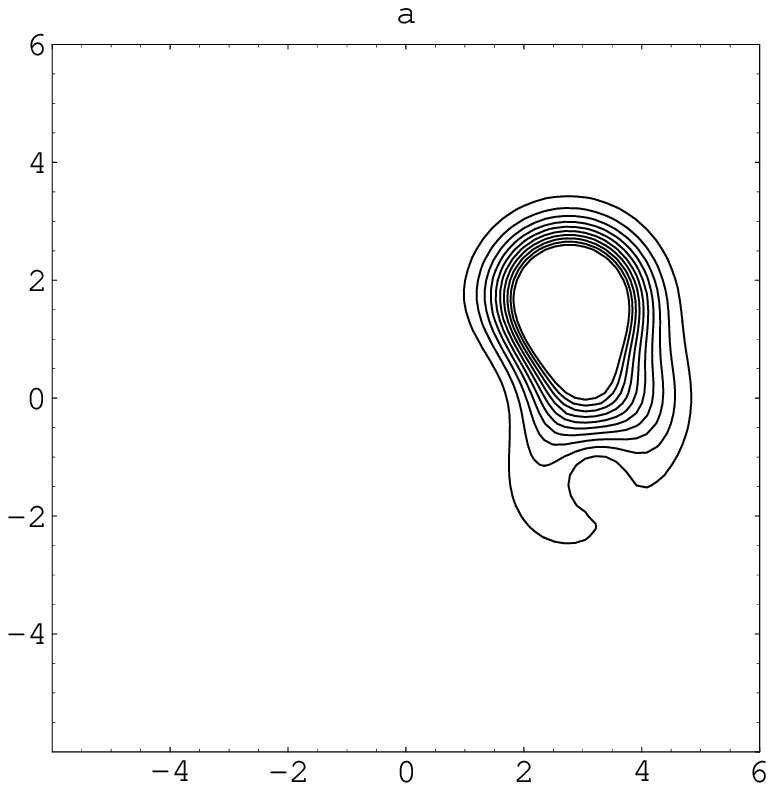}
\includegraphics[width=.3\linewidth]
{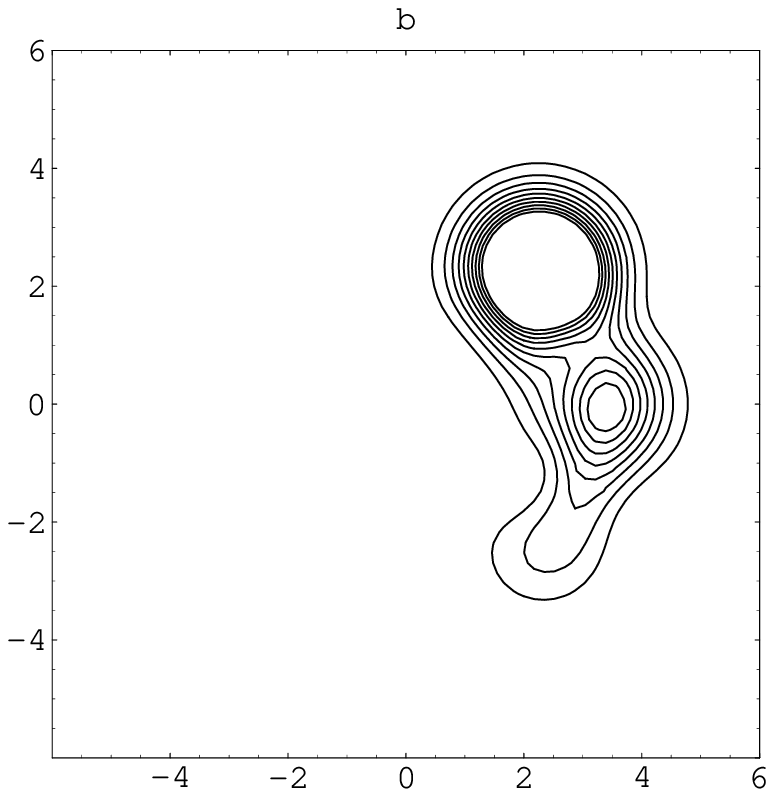}\\
\includegraphics[width=.3\linewidth]
{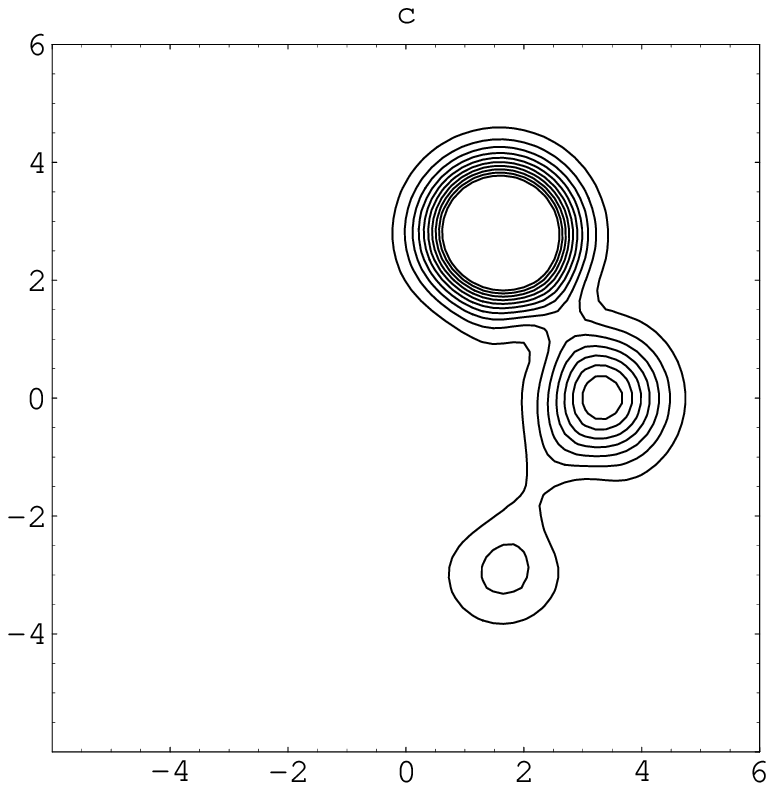}
\includegraphics[width=.3\linewidth]
{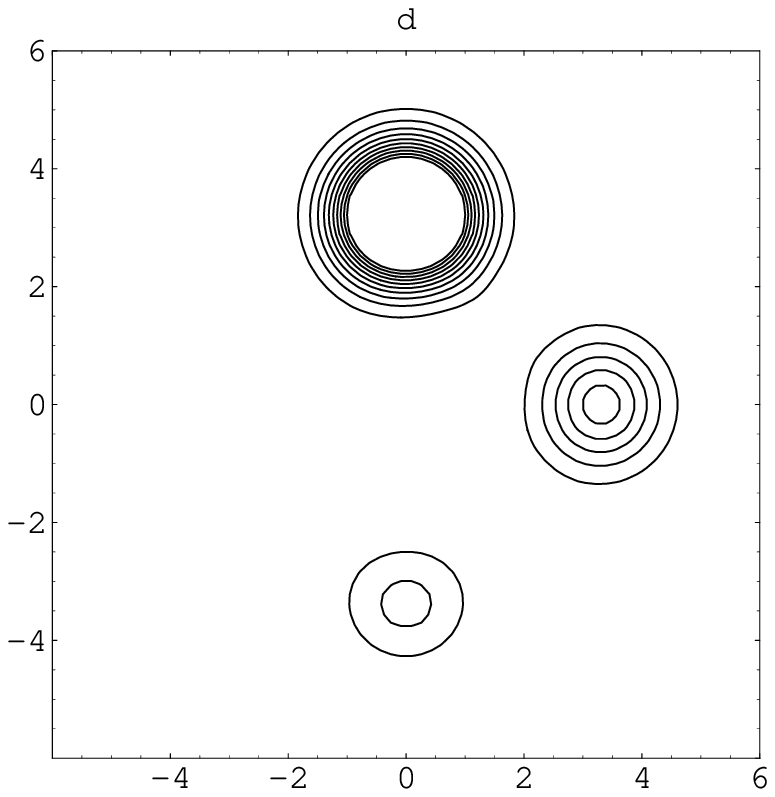}
\end{center}
\caption{The same as Fig. 7 but for $\chi/\lambda=0.5$ where (a) $\lambda t=\pi/6$ (b) $\lambda t=\pi/4$ (c) $\lambda t=\pi/3$ (d ) $\lambda t=\pi/2$}
\end{figure}
Here, we present a detailed analytical discussion about how the Husimi $Q$-function plays an important role in quantifying atom-atom entanglement. For this reason, we have pictured the numerical results obtained as contour plots in figures 7-9. The parameters was considered as follows: the average photon number $\bar n$ is kept fixed for all plots such as $\bar n=10$, the detuning parameter carries values such as $\delta=0.0,10.0$, and the Kerr parameter $\chi/\lambda$ varies as $\chi/\lambda=0.0, 0.2, 0.5, 1.0$, where the Rabi angle $\lambda t=5\pi/3$ is kept fixed for figures 7 and 8, while for figure 9 it changes as $\lambda t =\pi/6, \pi/4, \pi/3, \pi/2$, where the Kerr parameter $\chi/\lambda$ is half integer.\\
Form figure 7, by which we examine the effect of the Kerr parameter $\chi/\lambda$ at a fixed Rabi angle $\lambda t=5\pi/3$, we observe that, in the absence of the Kerr parameter, Fig. 7a, the Husimi $Q$-function composed of three components, two blobs centered at the points (0, $\pm4$) and a symmetric circular peak (Poisson band) centered at the point (4, 0) and strongly localized in the right half of the available phase space, i.e., the interaction region~\cite{THMAMAPO07}, once more, the squeezing effect on the number of photons can be observed from the figure. Experimental and theoretical studies showed that squeezing of matter wave fields (squeezed states) is closely related to entanglement~\cite{RSCNB03,RECL97,DSCZ00,SDCZ01,ATETO010,PMBLMH04,FIUNIO03}. Moreover, we observe that the Husimi distribution blobs delocalized almost around the right half of the phase space, i. e., the interaction region of phase space and have structures with almost positive density throughout the available phase space. It is worth to note that the  bifurcating of the blobs corresponds to the collapse. This can be clearly seen if we compare the numerical calculations for total populations and $Q$-function. It was shown that the bifurcation of the Husimi distribution, which is the signature of the formation of Schr{\"o}dinger cat states, implies correlation, and hence entanglement~\cite{VAOR95, ORPAKA95, JEOR94, MIMAWA00, MIWAIM01, SUGITA03, HIMCMI05}, see also Fig. 1a. As soon as the effect of the Kerr medium is considered, a clear difference in the figures shape appears, where the sensibility of the Husimi $Q$-function with the nonlinearity parameter $\chi/\lambda$ is very intersting. We observe that the Husimi distribution is taken gradually towards the non-interaction region, namely, left half of the available phase space, where we observe that the two blobs rotating in the complex coherent state parameter plane in the counterclockwise directions with the same speed where two overlapping regions appear with both terms are appreciably vanishing. But note also that the intersection areas with this prescription are not in the same position. Upon the increase of the Kerr parameter then one can observe that the overlapping becomes dominant and the Husime distribution are taken completely into the interaction region. As a consequence, strong entanglement degree between the two atoms can be built up at that time, see Figs. 1c and 7c. The conclusion will be extremely interesting when we look at the behaviors of both the concurrence $C$ and the Husimi distribution once the Kerr medium is taken into account, specially at the moment when the Kerr medium becomes strong at which entanglement diminishes and the Husimi distribution is strongly localized and taken into non-interaction region again, Figs. 1d and 7d. It is important to note that, localization of the Husimi distribution around a fixed point implies low degree of entanglement~\cite{SUGITA03, HIMCMI05}. 

\par Our observations will become clear when we look at the behaviors of both the concurrence $C$ and the Husimi distribution at the moment when detuning parameter possesses a value. The results showed in Figs. 2c, 3b and 8. A clear delocalization of the Husimi distribution but into the right half of phase space, the interaction region, corresponds to longer time interval of entanglement, Figs. 2c and 8a, while the opposite is true, Figs. 3b and 8b. An extremely interesting observations can be obtained when we consider various values of the Rabi angle $\lambda t$, while the nonlinear media is switched on. The results show strongly the role that the Hisimi distribution play in quantifying entanglement, where at all moments but not $\lambda t=\pi/2$, once the suitable Kerr parameter is chosen, stronger entanglement can be built up since the Husimi distribution is delocalized but completely spreads between two or three points into the right half of phase space, the interaction region, see also Fig. 1c.
\subsection{Injection of excited and ground atoms one by one}
\label{subsec:6.2}
In this case, for the states $\mid\begin{array}{l}
\vspace{-2ex}+\\
             +\\
\end{array}\rangle$ and $\mid\begin{array}{l}
\vspace{-2ex}-\\
             -\\
\end{array}\rangle$, $\rho_{F}$ reads
\begin{equation}
\label{eq91}
\rho_{F}=\biggl| S_{\begin{array}{l}
\vspace{-2ex}+\\
             +\\
\end{array}
}(t),n-1\biggr >\biggl< S_{\begin{array}{l}
\vspace{-2ex}+\\
             +\\
\end{array}
}(t),n-1\biggr| +\biggl| S_{\begin{array}{l}
\vspace{-2ex}-\\
             -\\
\end{array}
}(t),n+1\biggr >\biggl< S_{\begin{array}{l}
\vspace{-2ex}-\\
             -\\
\end{array}
}(t),n+1\biggr|.
\end{equation}
The $Q$-function is given by
\begin{equation}
\label{eq92}
Q=\frac{1}{\pi}\biggl(\biggl|\biggl<\alpha,\begin{array}{l}
\vspace{-2ex}+\\
             +\\
\end{array}\biggl|S_{\begin{array}{l}
\vspace{-2ex}+\\
             +\\
\end{array}},n-1\biggr>\biggr|^{2}+\biggl|\biggl<\alpha,\begin{array}{l}
\vspace{-2ex}-\\
             -\\
\end{array}\biggl|S_{\begin{array}{l}
\vspace{-2ex}-\\
             -\\
\end{array}},n+1\biggr>\biggr|^{2}\biggr),
\end{equation}
where 
\begin{equation}
\label{eq93}
\biggl<\alpha,\begin{array}{l}
\vspace{-2ex}+\\
             +\\
\end{array}\biggl| S_{\begin{array}{l}
\vspace{-2ex}+\\
             +\\
\end{array}
}(t),n-1\biggr >=e^{-(\mid\alpha\mid^{2}+\mid\alpha_{0}\mid^{2})/2}\sum_{n}\frac{(\alpha_{0}\alpha^{\ast})^{n}\alpha_{0}}{\sqrt{n!(n+1)!}}\Gamma_{1}(n+1,t)\Gamma_{2}(n+1,t),
\end{equation}
\begin{equation}
\label{eq94}
\biggl<\alpha,\begin{array}{l}
\vspace{-2ex}-\\
             -\\
\end{array}\biggl| S_{\begin{array}{l}
\vspace{-2ex}-\\
             -\\
\end{array}
}(t),n+1\biggr >=e^{-(\mid\alpha\mid^{2}+\mid\alpha_{0}\mid^{2})/2}\sum_{n}\frac{(\alpha_{0}\alpha^{\ast})^{n}\alpha^{\ast}}{\sqrt{n!(n+1)!}}\Gamma_{1}^{\ast}(n,t)\Gamma_{2}(n+1,t),
\end{equation}
\begin{figure}[tpbh]
\noindent
\begin{center}
\includegraphics[width=.3\linewidth]
{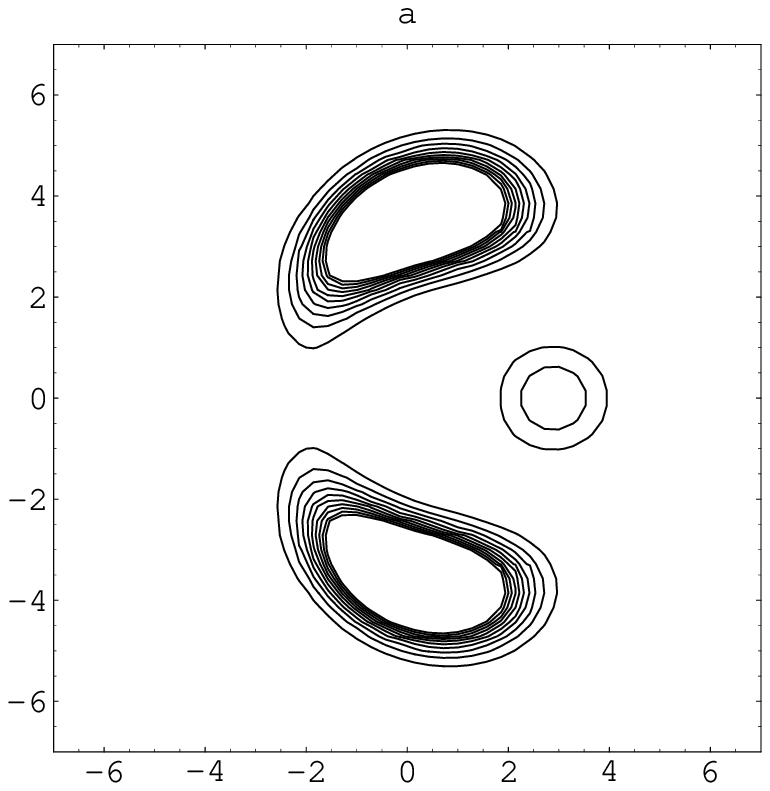}
\includegraphics[width=.3\linewidth]
{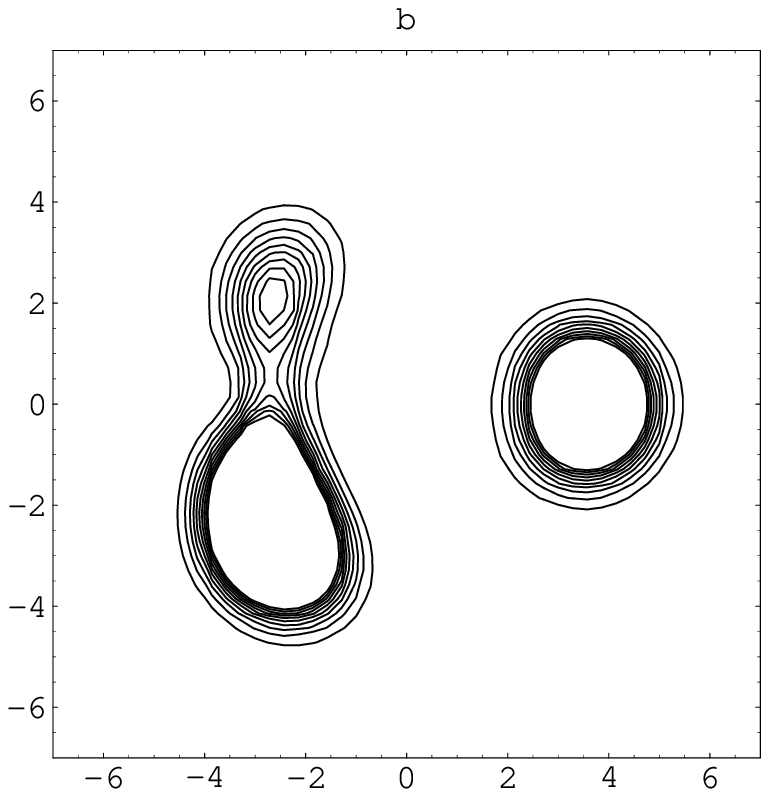}\\
\includegraphics[width=.3\linewidth]
{eeq3.eps}
\includegraphics[width=.3\linewidth]
{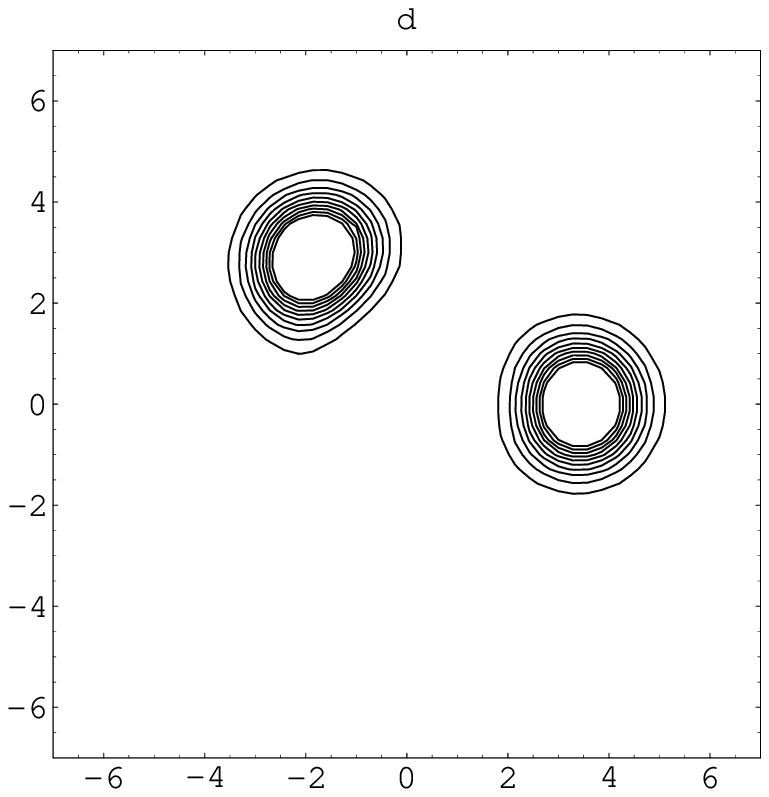}
\end{center}
\caption{Conour plots for Husimi $Q$-function  for $\bar{n}=10$, $\delta=0.0$, $\lambda t=5\pi/3$ where (a) $\chi/\lambda=0.0$ (b) $\chi/\lambda=0.2$ (c) $\chi/\lambda=0.5$ (d) $\chi/\lambda=1.0$}
\end{figure}
\begin{figure}[tpbh]
\noindent
\begin{center}
\includegraphics[width=.3\linewidth]
{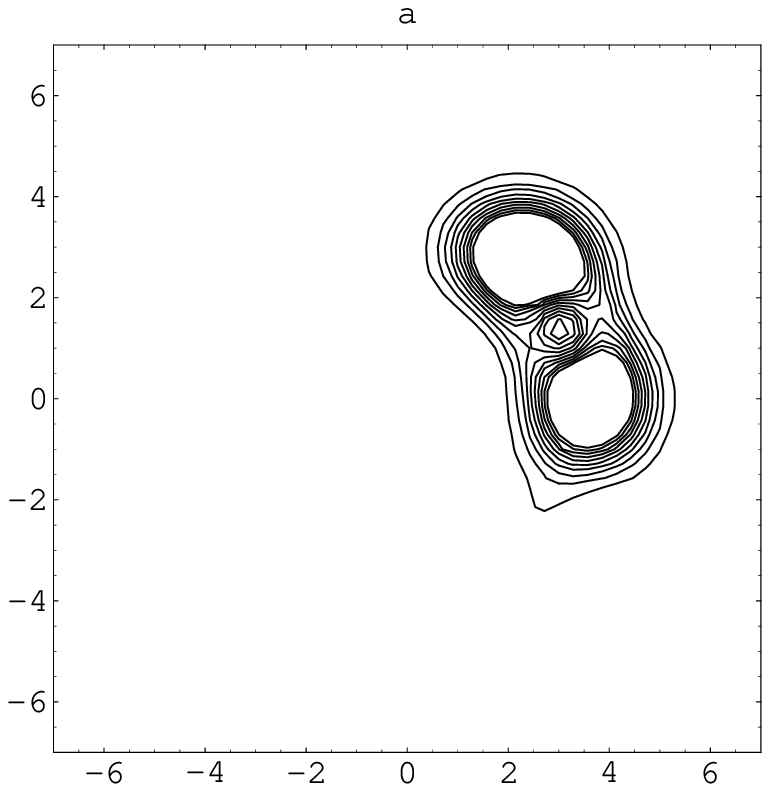}
\includegraphics[width=.3\linewidth]
{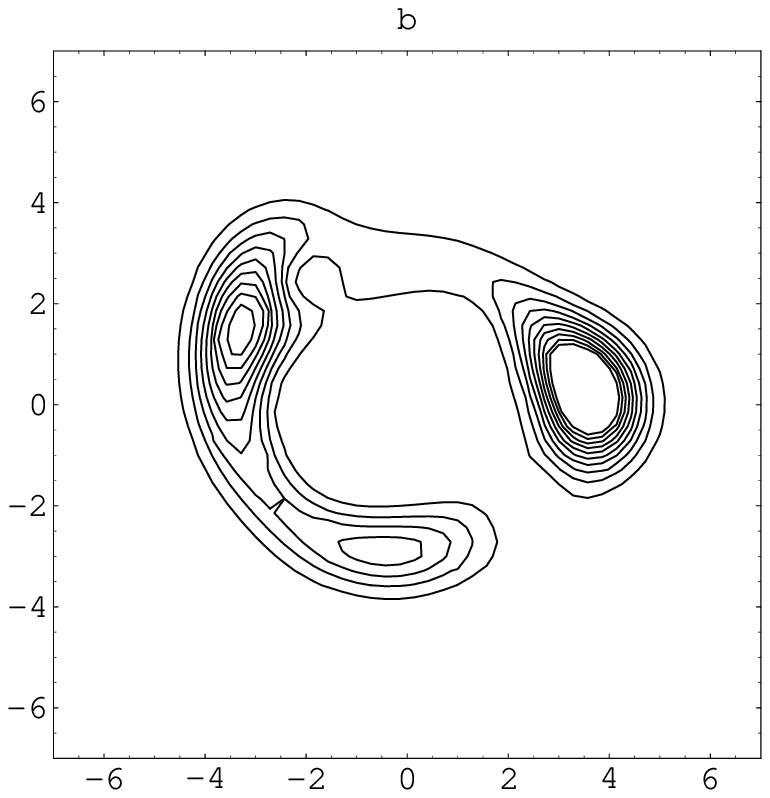}
\end{center}
\caption{The same as Fig. 10 but when $\delta=10.0$ and (a) $\chi/\lambda=0.0$ (b) $\chi/\lambda=1.0$}
\end{figure}
\begin{figure}[tpbh]
\noindent
\begin{center}
\includegraphics[width=.3\linewidth]
{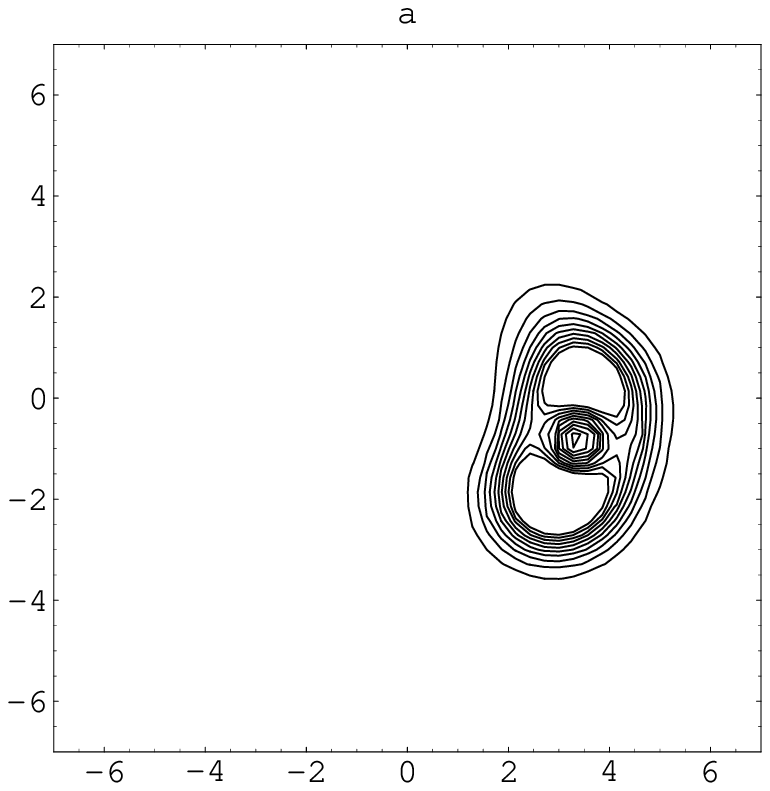}
\includegraphics[width=.3\linewidth]
{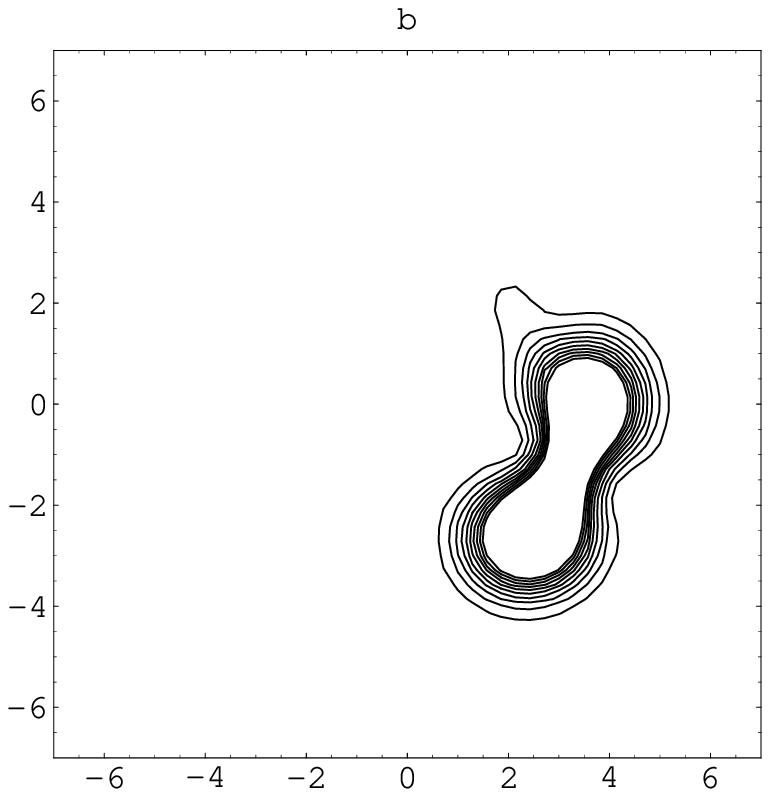}\\
\includegraphics[width=.3\linewidth]
{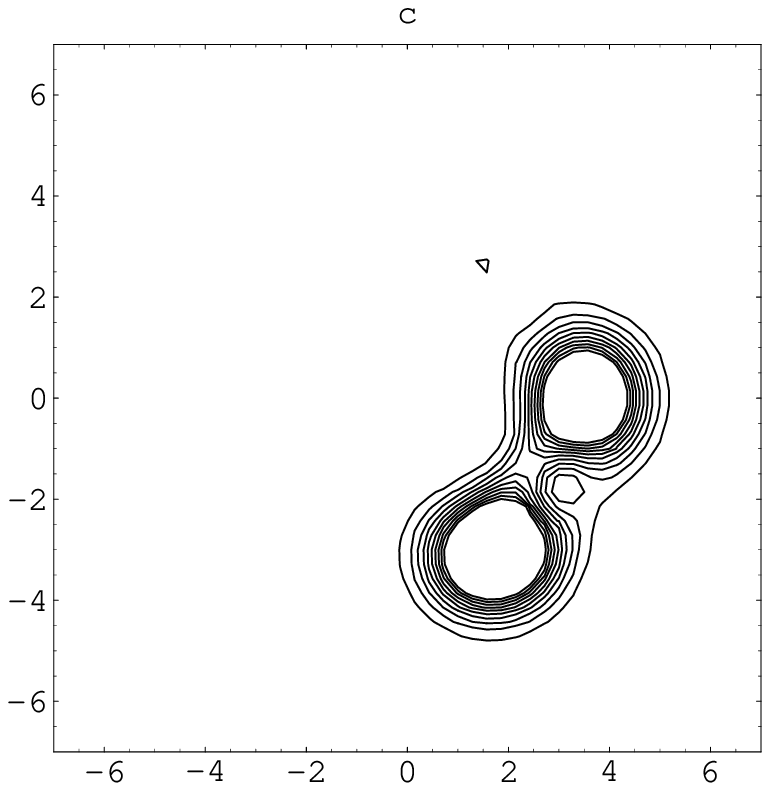}
\includegraphics[width=.3\linewidth]
{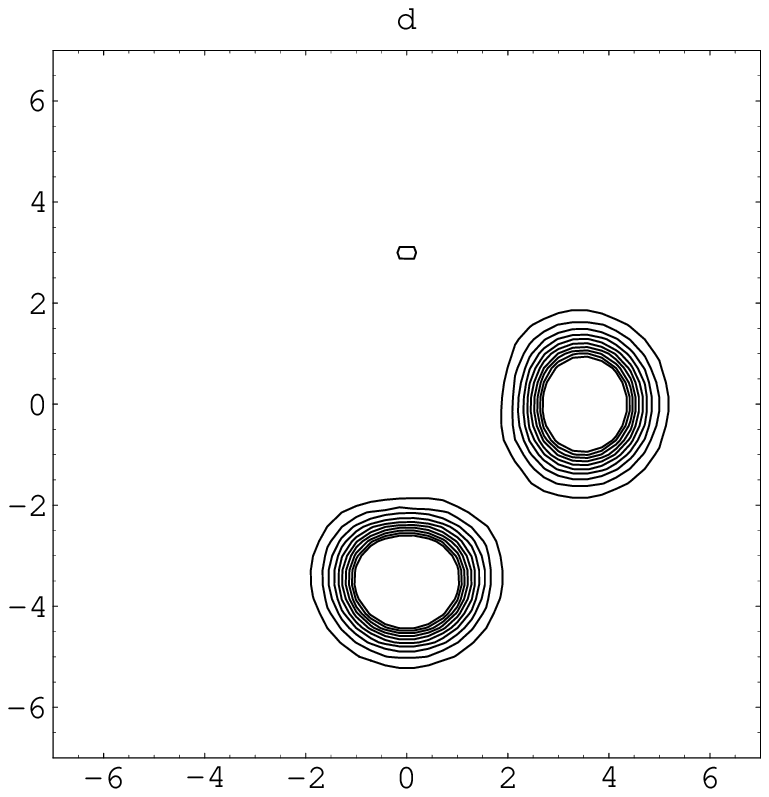}
\end{center}
\caption{The same as Fig. 10 but for $\chi/\lambda=0.5$ where (a) $\lambda t=\pi/6$ (b) $\lambda t=\pi/4$ (c) $\lambda t=\pi/3$ (d ) $\lambda t=\pi/2$}
\end{figure}
To obtain clearer insight about the efficiancy of the Husimi $Q$-function in quantifying entanglement, a comparative discusion will be important. Here we examine the behavior of the Husimi function under the same circumstances of parameters as in the previous section. This is illustrated in figures 10-12. Generally, the Husimi $Q$-function developed in manner similar to that observed previousley with slight change such that the localized Poisson band barely appear which means that nonclassical effect is not pronounced. In this case the entanglement between the two atoms doesn't live much more in opposite to the case of two successive excited atoms, where entanglement does appear at almost the whole time scale, see Figs. 1a, 4a, 7a, 10a. The case when we turn on the nonlinear media is also intersting. The Husimi distribution moves quickly towards the non-interaction region once the a weak Kerr medium is switched on, while a strong localized Poisson band appear clearly in the interaction region which reflects small probability of interaction and so lesser times of entanglement, see Figs. 4b, 10b. The suitable choice of the Kerr parameter push the Husimi distribution towards the interaction region again, which means stronger entanglement appears at longer times, Figs. 4c, 10c. There is no something interesting in case when the nonlinear media is strong, where almost the same behavior was seen before, for comparison see Figs. 1d, 4d, 7d, 10d. 
\par However, the effect of the detuning parameter $\delta$ is quite differente, specially in the presence of strong nonlinear media. In such a case, the Husimi distribution displays a strong delocalization and spreads between two points in the right half of the phase space in case of zero-value nonlinear media accompanied with a significant density throughout the available phase space. Moreover, overlapp takes place and the squeezing effect on the number of photons occurs can be observed. The strong Kerr media have intersting effect where the Husimi distribution shows ring-like shape around the center of the phase space but nevertheless have a significantly different densities throughout the available phase space, i.e., interaction and non-interaction regions. The effect of the matching choice the nonlinear media parameter $\chi/\lambda$ on the Husimi distribution at different values Rabi angle $\lambda t$ points out that enhancement of entanglement depends crucially on this suitable choice, see Figs. 4a and 12a-c.
\section{Conclusions}
We have considered the nature of the entanglement of output two successive atoms from a micromaser cavity for pure states input of the atoms when the cavity distribution is coherent. We have considered two different successive injections of the initial states of the atoms that traverse the cavity, namely, initially excited atoms and initially excited atom followed ground one. We have examined the effect of the frequency difference between the inter-atomic frequency and cavity frequency individually and in coexistence of nonlinear media on nonlocal atomic entanglement. We have also investigate the role of the Husimi $Q$-distribution that plays to give clearer insight about entanglement dynamics. Our conclusions are Summarized in the following:\\
(i) The physical nature of the interacting objects and the character of their mutual coupling control strongly the degree of quantum atomic entanglement, in other words, the interaction of a cavity field with two successive excited atoms plays more efficient role in producing atomic entanglement than that of interaction with two successive atoms traverse the cavity in dissimilar initial states.\\
(ii) The nonlinear medium plays an important role in producing atomic entanglement depending on the appropriate choice of its parameter. In this case an excellent periodical entanglement can be produced with increased maximum value comparing with the case when the nonlinear medium parameter is not appropriate or zero.\\
(iii) The extremely interesting is the role of the detuning parameter on producing nonlocal atomic entanglement. When the detuning parameter is considered to be considerably high, periodical, long lived  entanglement can be obtained.\\
(iv) Husimi $Q$-distribution plays a very clear and efficient role in quantifying nonlocal atomic entanglement, where bifurcating of Husimi $Q$-distribution in whole phase space and delocalization in the interaction region of the phase space corresponding to correlations between particles and as a consequence nolocal atomic entanglement.


\begin{thebibliography}{}
\bibitem{PWK04} N. A. Peters, T.-C. Wei, and P. G. Kwiat, \textit{Phys. Rev.} A \textbf{70}, (2004) 052309.
\bibitem{BBPS96} C. H. Bennett, H. J. Bernstein, S. Popescu, and B. Schumacher, \textit{Phys. Rev.} A\textbf{53}, (1996) 2046.
\bibitem{BDSW96} C. H. Bennett, D. P. DiVincenzo, J. A. Smolin, and W. K. Wootters,
\textit{Phys. Rev.} A \textbf{54}, (1996) 3824.
\bibitem{VP96} V. Vedral and M. B. Plenio, \textit{Phys. Rev.} A \textbf{57}, (1996) 1619.
\bibitem{V02} V. Vedral, \textit{Rev. Mod. Phys.} \textbf{74}, (2002) 197.
\bibitem{VT99} G. Vidal and R. Tarrach, \textit{Phys. Rev.} A \textbf{59}, (1999) 141.
\bibitem {VPRK95} V. Vedral, M. B. Pienio, M. A. Rippin and P. L. Knight, \textit{Phys. Rev.
Lett.} \textbf{78}, (1995) 2275.
\bibitem {W98} W. K. Wootters, \textit{Phys. Rev. Lett.} \textbf{80}, (1998) 2245.
%
\bibitem{GMN06} B. Ghosh, A. S. Majumdar and N. Nayak, \textit{Phys. Rev.} A \textbf{74}, (2006) 052315.
%
\bibitem {CZDKAW00} J. I. Cirac and P. Zoller, \textit{Phys. Rev.} A \textbf{50}, (1994) R2799; J. I. Cirac, P. Zoller, H. J. Kimble, and H. Mabuchi, \textit{Phys. Rev. Lett.} \textbf{78}, (1997) 3221; L.-M. Duan, M. D. Lukin, J. I. Cirac, and P. Zoller, Nature London \textbf{414}, (2001) 413; E. Solano, G. S. Agarwal, and H. Walther, \textit{Phys. Rev. Lett.} \textbf{90}, (2003) 027903; S. G. Clark and A. S. Parkins, \textit{ibid.} \textbf{90}, (2003) 047905; L.-M. Duan, B. Wang, and H. J. Kimble, \textit{Phys. Rev.} A \textbf{72}, (2005) 032333; G. S. Agarwal and K. T. Kapale, \textit{ibid.} \textbf{73}, (2006) 022315.
\bibitem{BGASMNN07} B. Ghosh, A. S. Majumdar and N. Nayak, \textit{Int. J. Quant. Inf.} \textbf{5}, (2007) 169.
\bibitem{PK491} S. J. D. Phoenix and P. L. Knight, \textit{Phys. Rev.} A \textbf{44}, (1991) 6023. 
\bibitem{PK691} S. J. D. Phoenix and P. L. Knight, \textit{Phys. Rev. Lett.} \textbf{66}, (1991) 2833.
\bibitem{KLMCK93} I. K. Kudryavtsev, A. Lambrecht, H. Moya-Cessa and P. L. 
knight, \textit{J. Mod. Opt.}, \textbf{40}, (1993) 1605.
\bibitem{ATETO109} M. S. Ateto, \textit{Int. J. Theor. Phys.} \textbf{48}, (2009) 545
\bibitem{Ateto07} M. S. Ateto, \textit{Int. J. Quant. Inf.}, \textbf{5(4)}, (2007) 535.
%
\bibitem{HMNWBRH979} E. Hagley, X. Maitre, G. Nogues, C. Wunderlich, M. Brune, J. M. Raimond, and S. Haroche, \textit{Phys. Rev. Lett.} \textbf{79}, (1997) 1; A. Rauschenbeutel, P. Bertet, S. Osnaghi, G. Nogues, M. Brune, J. M. Raimond, and S. Haroche, \textit{Phys. Rev.} A \textbf{64}, (2001) 050301 R; A. Auffeves, P. Maioli, T. Meunier, S. Gleyzes, G. Nogues, M. Brune, J. M. Raimond, and S. Haroche, \textit{Phys. Rev. Lett.} \textbf{91}, (2003) 230405; S. Nussmann, M. Hijlkema, B. Weber, F. Rohde, G. Rempe, and A. Kuhn, \textit{ibid.} \textbf{95}, (2005) 173602; D. N. Matsukevich, T. Chaneliere, S. D. Jenkins, S.-Y. Lan, T. A. B. Kennedy, and A. Kuzmich, \textit{ibid.} \textbf{96}, (2006) 030405.
%
%
\bibitem {KLAK02} M. S. Kim, J. Lee, D. Ahn and P. L. Knight, \textit{Phys. Rev.} A \textbf{65}, (2002) 040101(R); L. Zhou, H. S. Song and C. Li, \textit{J. Opt. B: Quantum
Semiclass. Opt.} \textbf{4}, (2002) 425.
%
\bibitem{BRAUN02} D. Braun, \textit{Phys. Rev. Lett.} \textbf{89}, (2002) 277901.
%
\bibitem{ATETO209} M. S. Ateto, \textit{Applied Mathematics $\&$ Information Sciences}, \textbf{3}, (2009) 41.
%
%
\bibitem{MLBEHW96} M. L$\ddot{o}$ffler, B.-G. Englert and H Walther, \textit{Appl. Phys.} B \textbf{63}, (1996) 511.
%
\bibitem{ASMNN01} A. S. Majumdar and N. Nayak, \textit{Phys. Rev.} A \textbf{64}, (2001) 013821.
%
\bibitem{RSWRWR991} G. Rempe, F. Schmidt-Kaler, and H. Walther, \textit{Phys. Rev. Lett.} \textbf{64}, (1990) 2783; G. Rempe, and H. Waklther, \textit{Phys. Rev.} A \textbf{42}, (1990) 1650; H. Paul, and Th. Richter, \textit{Optics Commun.} \textbf{85}, (1991) 508; J. D. Cresser, \textit{Phys. Rev.} A \textbf{46}, (1992) 5913.
\bibitem {M02} P. Masiak, \textit{Phys. Rev.} A \textbf{66}, (2002) 023804.
%
\bibitem{FIJAME86} P. Filipowicz, J. Javanainen, and P. Meystre, \textit{Phys. Rev.} A \textbf{34}, (1986) 3077.
%

\bibitem{PK88} S. J. D. Phoenix and P. L. Knight, \textit{Ann. Phys.} \textbf{186}, (1988) 381.
%
\bibitem{PB93} S. J. D. Phoenix and S. M. Barentt, \textit{J. Mod. Opt.}, \textbf{40}, (1993) 979.
%
\bibitem{SUGITA03} A. Sugita, \textit{J.Phys. A: Math. Gen.} \textbf{36}, (2003) 9081.
%
\bibitem{FALI95} M.-F. Fang and H.-E. Liu, \textit{Phys. Lett. A} \textbf{200}, (1995) 250.
%
\bibitem{ATETO10} M. S. Ateto, \textit{Int. J. Theor. Phys.} \textbf{49}, (2010) 276.
%
\bibitem {WLG06} C.-Z. Wang, C.-X. Li and G.-C. Guo, \textit{Eur. Phys. J.} \textbf{37}, (2006) 267.
%
%
\bibitem{JOPU92} A. Joshi and R. R. Puri, \textit{Phys. Rev.} \textbf{A 45}, (1992) 5056.
\bibitem{Hillaer91} M. Hillery, \textit{Phys. Rev.} \textbf{A 44}, (1991) 4578.
\bibitem{CHCOWO92} A.N. Chaba, M.J. Collett, D.F. Walls, \textit{Phys. Rev.} \textbf{A 46}, (1992) 1499.
\bibitem{ZHGCLM00} R. Zambrini, M. Hoyuelos, A. Gatti, P. Colet, L. Lugiato, M.S. Miguel, \textit{Phys. Rev.} \textbf{A 62}, (2000) 063801.
\bibitem{Gerry99} C.C. Gerry, \textit{Phys. Rev.} \textbf{A 59}, (1999) 4095.
\bibitem{PKH03} M. Paternostro, M.S. Kim, B.S. Ham, \textit{Phys. Rev.} \textbf{A 67}, (2003) 023811.
\bibitem{PACH00} J. Pachos, M. Chountasis, \textit{Phys. Rev.} \textbf{A 62}, (2000) 052318.
\bibitem{VFT00} D. Vitali, M. Fortunato, P. Tombesi, \textit{Phys. Rev. Lett.} \textbf{85}, (2000) 445.
%
\bibitem {DGMN04} A. Datta, B. Ghosh, A. S. Majumdar and N. Nayak, \textit{Europhys. Lett.} \textbf{67}, (2004) 934.
\bibitem{LEW96} M. L$\ddot{o}$ffler, B.-G. Englert and H Walther, \textit{Appl. Phys.} \textbf {B 63}, (1996) 511.
\bibitem{AMNN01} A. S. Majumdar and N. Nayak, \textit{Phys. Rev.} \textbf{A 64}, (2001) 013821.
%
\bibitem{EGSWH96} B.-G. Englert T. Gantsog, A. Schenzle, C. Wagner and H Walther, \textit{Phys. Rev.} \textbf{A 53}, (1996) 4386.
\bibitem{MHNWBTH97} X. Maˆitre, E. Hagley, G. Nogues, C. Wunderlich, P. Goy, M. Brune, J. M. Taimond and S. Haroche, \textit{Phys. Rev. Lett.} \textbf{79}, (1997) 769.
\bibitem{ELSW002} B.-G. Englert, P. Lougovski, E. Solano and H. Walther, e-print: quant-ph/0209128v1 24Sep 2002.
\bibitem{HHH95} T. Horodecki, P. Horodecki and M. Horodecki, \textit{Phys. Lett.} \textbf{A 200}, (1995) 340.
\bibitem{EWIGNER32} E. Wigner, \textit{Phys. Rev.} \textbf{40}, (1932) 749.
\bibitem{ZWIGNer32} Z. Wigner, \textit{Phys. Chem.} \textbf {B 19}, (1932) 203.
\bibitem{KAGL69} K. E. Cahill and R. J. Glauber, \textit{Phys. Rev.} \textbf{177}, (1969) 1882.
\bibitem{HICOSCWG84} M. Hillary, R. F. O'Conell, M. O. Scully and E. wigner, \textit{Phys. Rep.} \textbf {106}, (1984) 121.
\bibitem{BOTAWA98} E. L. Bolda, S. M. Tan and D. Walls, \textit{Phys. Rev.} \textbf{A 57}, (1998) 4686.
\bibitem{MANTOM97} S. Mancini and P. Tombesi, \textit{Europhys. Lett.} \textbf{40}, (1997) 352.
\bibitem{WEHRL79} A. Wehrl, \textit{Rep. Math. Phys.} \textbf{16}, (1979) 353.
\bibitem{PEKRPELUSZ86} V. Pe$\breve{r}$inov$\acute{a}$, J. K$\breve{r}$epelka, J. Pe$\breve{r}$ina, J. Luk$\breve{s}$ and P. Szlachetkta, \textit{Opt. Acta} \textbf{33}, (1986) 15.
\bibitem{FAGU06} Hong-yi Fan and Qin Guo, quant-ph/0611206v1 (2006).
\bibitem{CAALCARA09} C. P.-Campos, J. R. G.-Alonso, O. Casta$\tilde{n}$os and R. L.-Pe$\tilde{n}$a, cond-mat.guant-gas/0910.3256v1 (2009).
\bibitem{HUFAN09} Li-yun Hu and Hong-yi Fan, quant-ph/0911.0125v1, (2009).
\bibitem{MIMAWA00}A. Miranowicz,J. Bajer, M . R. B. Wahiddin and N. Imoto, \textit{J. Phys A: Math. Gen.} \textbf{34}, (2001) 3887.
\bibitem{MIWAIM01}A. Miranowicz, H. Matsueda and M. R. B. Wahiddin, \textit{J. Phys. A: Math. Gen.} \textbf {33}, (2000) 5159.
\bibitem{BERETA84} G. P. Beretta, \textit{J. Math. Phys.} \textbf{25}, (1984) 1507.
\bibitem{HUSIMI40} K. Husimi, \textit{Proc. Phys. Math. Soc. Japan} \textbf{22}, (1940) 264.
\bibitem{FuSOLO001} H. Fu and A. I. Solomom, \textit{J. Mod. Opt.} \textbf{49}, (2002) 259.
%
\bibitem{THMAMAPO07} Thomas Mainiero and   Mason A. Porter, arXiv:nlin/0702025v3 [nlin.CD] 20 Sep (2007).
%
\bibitem{RSCNB03} J. Rogel-Salazar, S. Choi, G.H.C. New, and K. Burnett, \textit{Phys. Rev.} \textbf{A 65}, (2002) 023601.
\bibitem{RECL97} W. P. Reinhardt and C. W. Clark, \textit{J. Phys. B.} \textbf{30}, (1997) L785.
\bibitem{DSCZ00} L. M. Duan, A. Sørensen, J. I. Cirac, P. Zoller, \textit{Phys. Rev. Lett.} \textbf{85},  (2000) 3991.
\bibitem{SDCZ01} A. Sørensen, L.M. Duan , J. I. Cirac, P. Zoller, \textit{Nature} \textbf{409}, (2001) 63.
\bibitem{ATETO010} M. S. Ateto, , arXiv: 0911.4240v2 [quant-ph], accepted in \textit{Int. J. Quant. Inf.}.
\bibitem{PMBLMH04} Amy Peng, Mattias Johnsson, W. P. Bowen, P. K. Lam, H. -A. Bachor and J. J. Hope, \textit{Phys. Rev.} \textbf{A 71}, (2005) 033809.
\bibitem{FIUNIO03}  Jarom$\acute{i}$r Fiur$\acute{a}\breve{s}$ek, and Nicolas J. Cerf, \textit{Phys. Rev. Lett.} \textbf{93}, (2004) 063601.
%
\bibitem{VAOR95} J. A. Vaccaro and A. Orlowski, \textit{Phys. Rev.} \textbf{A 51}, (1995) 4172.
\bibitem{ORPAKA95} A. Orlowski, H. Paul and G. Kastelewicz, \textit{Phys. Rev.} \textbf{A 52}, (1995) 1621.
\bibitem{JEOR94} I. Jex and A. Orlowski, \textit{J. Mod. Opt.} \textbf{41}, (1994) 2301.
\bibitem{HIMCMI05} Andrew P. Hines, Ross H. McKenzie, and G.J. Milburn, arXiv:quant-ph/0308165v4 28 Feb (2005)
%
%
\end{thebibliography}
\end{document}